\documentclass[11pt]{article}

\newcommand{\ii}{\mathrm{i}}

\newcommand{\al}{\alpha'}

\usepackage{tikz} 
\usetikzlibrary{decorations.markings} 
\usetikzlibrary{decorations.text} 
\usetikzlibrary{positioning} 
\usetikzlibrary{backgrounds} 
\usetikzlibrary{fit} 
\usetikzlibrary{calc} 
\usetikzlibrary{shapes} 

\tikzset{dot/.style={draw,circle,inner sep=.7pt,fill,node
    distance=1cm}} 
\tikzset{dot1/.style={draw,circle,inner sep=.7pt,fill}} 
\tikzset{triangle/.style={draw,regular polygon, regular polygon
    sides=3}} 
\tikzset{->-/.style={decoration={
  markings,
  mark=at position .5 with {\arrow{>}}},postaction={decorate}}} 
\tikzset{-<-/.style={decoration={ 
  markings,
  mark=at position .5 with {\arrow{<}}},postaction={decorate}}}

\usepackage{jheppub}
\usepackage{graphicx}

\title{Perturbative quantum gravity in double field theory}
\author{Rutger H. Boels}\emailAdd{rutger.boels@desy.de}
\author{and Christoph Horst}\emailAdd{mail@christoph-horst.de}
\affiliation{II. Institut f\"ur Theoretische Physik, Universit\"at Hamburg\\ Luruper Chaussee 149, D- 22761 Hamburg, Germany }

\abstract{We study perturbative general relativity with a two-form and a dilaton using the double field theory formulation which features explicit index factorisation at the Lagrangian level. Explicit checks to known tree level results are performed. In a natural covariant gauge a ghost-like scalar which contributes even at tree level is shown to decouple consistently as required by perturbative unitarity. In addition, a lightcone gauge is explored which bypasses the problem altogether. Using this gauge to study BCFW on-shell recursion, we can show that most of the D-dimensional tree level S-matrix of the theory, including all pure graviton scattering amplitudes, is reproduced by the double field theory. More generally, we argue that the integrand may be reconstructed from its single cuts and provide limited evidence for off-shell cancellations in the Feynman graphs. As a straightforward application of the developed technology double field theory-like expressions for four field string corrections are derived.}

\begin{document}
\maketitle

\section{Introduction}
As far as known nature encompasses four fundamental forces. Three of these, the electromagnetic as well as the strong and weak nuclear forces are contained within the standard model of particle physics. This model is formulated as a Lagrangian perturbative quantum field theory, with a strong focus on symmetries such as those of special relativity. The fourth force of nature is gravity. As a theory of nature this is encapsulated in the theory of general relativity. Although this can be formulated as a classical Lagrangian field theory, perturbative quantisation of this theory leads to difficulties, usually captured in the phrase that "the theory is not renormalisable". Hence perturbative quantisation does not seem to lead to a consistent effective quantum theory at experiment-accessible length scales: degrees of freedom at arbitrarily high energy scales which could influence low energy processes by the rules of quantum mechanics cannot be shown to decouple. Technically, this is seen by inspecting ultra-violet divergences arising in loop diagrams within the Feynman graph approach to perturbation theory. 

However, the Feynman graph approach to perturbative Einstein gravity is technically exceedingly complicated. Starting with the Einstein Hilbert Lagrangian,
\begin{equation}
\mathcal{S}_{EH} = \int d\!x^D \sqrt{-g} R(g)
\end{equation}
one expands the metric field ($g$) around a flat background ($G$), 
\begin{equation}
g = G + h
\end{equation}
After choosing an appropriate gauge, graviton scattering amplitudes may be computed. The obtained expressions are generically a complicated mess. This is the result of the breaking of manifest local Lorentz invariance which is restored only in the final expression. Although the problem of calculational complexity can be pushed back by roughly a loop order using the background formalism, it is endemic to off-shell approaches to perturbative quantum gravity. 

The textbook way of approaching ultra-violet divergences in loop diagrams is therefore an estimate based on inspection of individual terms in the perturbative expression. Under the assumption there is no cancellation between terms, this gives the overall divergence. Any cancellation would naturally be associated to a symmetry. The archetypical example is the computation in \cite{'tHooft:1974bx} of the one-loop UV-divergent terms in pure Einstein gravity. They found the sum cancels as a consequence of diffeomorphism invariance: there are no diffeomorphism invariant terms in the Lagrangian which cannot be written as total derivatives.  Terms of this type are loosely referred to as counter-terms. For a long period the consensus view has been however that Einstein gravity in four dimensions is intrinsically non-renormalisable and that incurable divergences will set in. At two loops this has been verified explicitly \cite{Goroff:1985th} \cite{vandeVen:1991gw}, see \cite{Bern:2015xsa} for a recent discussion. Adding Poincar\'e supersymmetry was originally expected to improve the UV-behaviour to a divergence at three loops. 

Starting with \cite{Bern:2007hh}, there has been a remarkable shift in the canonical point of view for UV-divergences in especially maximally supersymmetric gravity theories. The generally accepted view, see for instance \cite{Beisert:2010jx} for a particularly clear approach, is that these most supersymmetric theories potentially diverges at seven loops in four dimensions (and at five loops in five dimensions). Various related results for less supersymmetric theories of gravity have been worked out. Even more interesting than these explicit results is the structure that drives these cancellations dubbed ``color-kinematic duality'' \cite{Bern:2008qj, Bern:2010ue}. 

Color-kinematic duality is a precise statement about the structure of the gravity integrand and its relations to that of a Yang-Mills theory. This had appeared already before in another guise which is of interest for the current paper. For free fields it is obvious that spin-two field occur within the tensor product of two spin-one fields. In string theory, it is known \cite{Kawai:1985xq} that this observation stretches to the full tree-level S-matrix. Basically, the left and right-moving modes of the string make up a `left' and a `right' open string contribution to the worldsheet correlation function. Most of the work is in showing how to disentangle the integrations over the closed string tree level worldsheet into two integrations over the left and right contributions and rewriting the result as sum over products of tree level open string amplitudes. In the limit that the string tension becomes very large compared to the typical scale of the momentum invariants, roughly $\alpha' \rightarrow 0$, the string S-matrix reduces to the S-matrix of Einstein gravity, coupled to an anti-symmetric two-tensor and a dilaton (the other two fields in the tensor product of two spin-one fields). The latter theory is sometimes referred to as `$\mathcal{N}=0$ supergravity'. The S-matrix of this theory inherits its double copy structure from the string theory. Understanding this factorisation property of the tree-level S-matrix from a purely field theory point of view has long been lacking. Color-kinematic duality is one way of making the factorisation manifest on a technical level in field theory. 

However, color-kinematic duality in its current form will only ever confirm the presence or absence of divergences at a specific, fixed order amplitude (see \cite{Boels:2010nw} for an attempt to go beyond). Even if a divergence is found, this does not necessarily mean the theory is not renormalisable: there could be a symmetry relating all counter-terms. See \cite{Bork:2015zaa} for speculation in this direction for gauge theory. A second, orthogonal and so far less influential way an understanding of a factorisation property in Einstein gravity has been achieved is through the action, see \cite{Bern:1999ji} and especially \cite{Hohm:2011dz} (see also \cite{Hohm:2010xe}). In the approach of \cite{Hohm:2011dz} general relativity is rewritten to incorporate string theory T-duality at the field theory level. The result is known as `double field theory' (DFT), see e.g. \cite{Hohm:2013bwa} for a review.  Potentially, this is a much more powerful way of approaching the question of UV-divergences as actions are usually simpler to handle than explicit scattering amplitudes. In this paper we push development of the double-field theory approach to understanding perturbative gravity further towards maturity. 

This paper is organised as follows: in section \ref{sec:rev} we establish our notation and introduce the double field theory formulation of  `$\mathcal{N}=0$ supergravity'. This is followed by a discussion of the Feynman graph perturbation theory in a covariant gauge in section \ref{sec:cov}. Of special importance is the analysis of unitarity in this gauge, which is not manifest. Several explicit calculations are presented. In the next section, section \ref{sec:lightcone}, we introduce lightcone gauge and work out the propagator in this gauge. Next, we use this to analyse large BCFW shift behaviour. Although our analysis does not reproduce the full known large shift behaviour, enough is obtained to show that most of the $\mathcal{N}=0$ tree level S-matrix (including all scattering amplitudes involving only gravitons) is reproduced by the double field theory perturbation theory. Some interesting patterns of cancellations are pointed out which hold for the gravity integrand. A discussion and conclusion sections ends the main presentation. Of independent interest may be appendix \ref{app:lightcone}, which also includes a computation of the Einstein-Hilbert lightcone gauge propagator, a result we have been unable to locate in the literature. 

A Mathematica notebook available at \href{http://www.desy.de/~boels}{this url} contains many explicit details of the computations reported here.

\section{Review of double field theory in a flat background}\label{sec:rev}
In this section the double field theory Lagrangian describing metric perturbations around a flat background is reviewed, roughly following \cite{Hohm:2011dz} (see also \cite{Hohm:2010xe}). The low-energy effective action of closed bosonic string theory in $D$-dimensional spacetime is 
\begin{equation}\label{lowenergyclosedstringaction}
  S = \int dx \sqrt{g} e^{-2\phi} \left (R+4\,\partial_i\phi \partial^i\phi -
    \frac{1}{12} H^2 \right)\, ,
\end{equation}
where $R$ is the Ricci-scalar of the metric $g_{ij}$ with ``mostly
plus'' signature, $H_{ijk}=3\partial_{[i} b_{jk]}$ is the 3-form field
strength of the 2-form $b_{ij}$, and $\phi$ is the dilaton
field. Moreover, here one abbreviates
$H^2=g^{ik}g^{jl}g^{pq}H_{ijp}H_{klq}$. In principle, this action could
be used to compute (tree-level) scattering amplitudes for quantum
excitations around a constant flat background,
\begin{equation}
  g_{ij}(x)=G_{ij}+g^{\text{(fluc.)}}_{ij}(x)\, , \qquad
  b_{ij}(x)=B_{ij}+b^{\text{(fluc.)}}_{ij}(x)\, ,\qquad\phi(x)=\langle
  \phi\rangle +
  \phi^{\text{(fluc.)}}(x)\, ,
\end{equation}
where $G_{ij}$ is the $D$-dimensional background Minkowski
metric again with ``mostly plus'' signature and $B_{ij},\langle
  \phi\rangle$ are the
constant backgrounds of the $B$-field and dilaton,
respectively.\footnote{Note that without loss of generality constant
  backgrounds $B_{ij}$ and $\langle \phi\rangle$ could be set to
  zero.} The derivation of Feynman rules in gravity theories such as 
\eqref{lowenergyclosedstringaction} is almost arbitrarily laborious due to an
infinite number of interaction terms in the Lagrangian. At the same
time, unlike the $S$-matrix that is protected by what is sometimes
called the ``equivalence theorem'' (stated in, e.g.  \cite{vandeVen:1991gw}
and reference [25] therein), Feynman rules and their resulting
individual Feynman diagrams do not have an intrinsic meaning for they
change with field redefinitions. As a consequence, a better choice of
field redefinitions can possibly simplify the computation of Feynman
rules and Feynman diagrams. In what follows we will summarise the
field redefinitions that lead to such a simplification for
\eqref{lowenergyclosedstringaction} and also circumvent the need to
explicitly expand the density factor $\sqrt{g}$ in $D$ dimensions. 

In \cite{Hohm:2011dz} it has been pointed out that another equivalent
double field theoretical formulation of
\eqref{lowenergyclosedstringaction} exists with a certain index 
factorisation property that is guaranteed by $O(D,D)$ ``T-duality'':
After a (non-linear) field redefinition tensor fluctuations are
described by a field $e_{ij}$ where $i$ ($j$) is called a left-index
(right-index), respectively, and the Lagrangian does not have terms
with mixed left-right index contractions. The DFT extension 
\cite{Hull:2009mi,Hohm:2010jy} of \eqref{lowenergyclosedstringaction}
is given in terms of objects $\mathcal{E}_{ij}=g_{ij}+b_{ij}$ and a 
density $d$ with $\sqrt{g}e^{-2\phi} = e^{-2d}$ that both depend on
the ``doubled spacetime coordinates'' $(\tilde{x},x)$. Its action reads
\begin{align}
\label{LDFT}
  S=\int dxd\tilde{x}\, e^{-2d} \Big [&-\frac{1}{4}g^{ik} g^{jl}
    \mathcal{D}^p \mathcal{E}_{kl} \mathcal{D}_p \mathcal{E}_{ij} +
    \frac{1}{4}g^{kl} \left (\mathcal{D}^j \mathcal{E}_{ik}
      \mathcal{D}^i \mathcal{E}_{jl} + \bar{\mathcal{D}}^j
      \mathcal{E}_{ki} \bar{\mathcal{D}}^i \mathcal{E}_{lj} \right)
  \nonumber\\
  & + \left (\mathcal{D}^id\, \bar{\mathcal{D}}^j \mathcal{E}_{ij} +
    \bar{\mathcal{D}}^id\, \mathcal{D}^j \mathcal{E}_{ji}\right)
  +4\mathcal{D}^id\,\mathcal{D}_i d \Big]\, .
\end{align}
As shown in \cite{Hohm:2010jy}, the DFT Lagrangian \eqref{LDFT} is an
extension of \eqref{lowenergyclosedstringaction} in that setting all
fields independent of $\tilde{x}^i$, i.e.\ $\tilde{\partial}=0$, the
two Lagrangians are equivalent (up to a total derivative which in
perturbation theory around a flat background one may safely  
discard). For fluctuations $e_{ij}$ defined as follows,
\begin{equation}
\label{definitionofe}
  \mathcal{E}_{ij} = G_{ij} + B_{ij} + {\left(\left(1-\frac{1}{2} e\,
      G^{-1}\right)^{-1} \right)_i}^k e_{kj}
  \, ,
\end{equation}
the DFT Lagrangian exhibits the left-right index
factorisation\cite{Hohm:2011dz}. Note that the factorisation at the
level of the Lagrangian descends to a factorisation at the level of
Feynman rules. The existence of such factorised Feynman rules fits
into the KLT \cite{Kawai:1985xq} picture where tree-level closed
string amplitudes factorize into (sums of) products of two tree-level
open string amplitudes. In the limit $\alpha'\rightarrow 0$ this
decomposition descends to one in which amplitudes of
\eqref{lowenergyclosedstringaction} are given in terms of products of
two color-ordered amplitudes associated to a ``left'' and ``right''
Yang-Mills gauge theory, respectively. However, while such relations
can be explicitly checked for low leg numbers it turns out that the
KLT decomposition is not manifest at the level of the double field theory Lagrangian despite the index factorisation. 

Subsequently, rather than using the DFT action \eqref{LDFT} and the
non-linear field redefinition \eqref{definitionofe}, on a technical
level it is even more convenient to use another equivalent action
written in terms of Siegel's frame fields \cite{Siegel:1993th} (his
formalism is nicely reviewed in \cite{Hohm:2011dz}). This is due to
the fact that in expanding \eqref{LDFT} to a given order in $e_{ij}$
at intermediate steps one has to deal with interaction terms that
violate the left-right index factorisation although eventually such
terms all cancel out. In contrast, it is by construction that Siegel's
action has manifest index factorisation. Siegel's Lagrangian reads  
\begin{align}
\label{Lsiegel}
  \frac{1}{4}\mathcal{L} =& -\frac{1}{2}\Phi^2 \mathcal{G}^{ab}
  \mathcal{G}^{\bar{c}\bar{d}} \Big (\mathcal{G}^{cd} {e_a}^M \nabla_c
  e_{\bar{c}M} {e_b}^N \nabla_d e_{\bar{d}N} - \mathcal{G}^{cd}
  {e_{\bar{c}}}^M \nabla_a e_{cM} {e_{\bar{d}}}^N \nabla_d
  e_{bN}\nonumber\\
  &\hspace{2.4cm}+\mathcal{G}^{\bar{a}\bar{b}} {e_a}^M \nabla_{\bar{a}} e_{\bar{c}M}
  {e_b}^N \nabla_{\bar{d}} e_{\bar{b}N} \Big )\nonumber\\
  &+\Phi \mathcal{G}^{ab}\mathcal{G}^{\bar{c}\bar{d}} \left ({e_a}^M
    \nabla_{\bar{c}} e_{\bar{d}M} \nabla_b \Phi - {e_{\bar{c}}}^M
    \nabla_a e_{bM} \nabla_{\bar{d}}\Phi \right) - 2\mathcal{G}^{ab}
  \nabla_a \Phi \nabla_b\Phi \, ,
\end{align}
where
\begin{align}
  {e_a}^M\nabla_b e_{\bar{c}M} & = D_b h_{a\bar{c}} - h_{b\bar{d}}
  D^{\bar{d}}h_{a\bar{c}}, & \nabla_a \Phi = D_a\varphi -
  h_{a\bar{b}} D^{\bar{b}} \varphi,\nonumber\\
  {e_{\bar{c}}}^M \nabla_a e_{bM} & =
  -D_ah_{b\bar{c}}+ h_{a\bar{d}} D^{\bar{d}} h_{b\bar{c}}, & \nabla_{\bar{a}} \Phi= D_{\bar{a}}\varphi +h_{b\bar{a}} D^b
  \varphi\, ,\nonumber\\
  {e_a}^M \nabla_{\bar{b}} e_{\bar{c}M} & = D_{\bar{b}} h_{a\bar{c}} +
  h_{d\bar{b}} D^d h_{a\bar{c}}\, ,&
\end{align}
are given in terms of fields $h_{a\bar{b}}$ and
$\varphi$. Furthermore, one has $\Phi=e^{-d} = 1 + \varphi$ and
tangent space metrics
\begin{align}
  \mathcal{G}_{ab} = \langle \mathcal{G}_{ab} \rangle +
  {h_a}^{\bar{c}} h_{b\bar{c}},\qquad \mathcal{G}_{\bar{a}\bar{b}} =
  \langle \mathcal{G}_{\bar{a}\bar{b}} \rangle + {h^c}_{\bar{a}}
  h_{c\bar{b}}\, , 
\end{align} 
whose inverses are given by
\begin{align}
  \mathcal{G}^{-1} = \langle \mathcal{G} \rangle^{-1}
  \sum_{n=0}^{\infty} (-1)^n \left ( h\langle \bar{\mathcal{G}}
    \rangle h^T \langle \mathcal{G} \rangle^{-1} \right )^n
\end{align}
and similarly for $\bar{\mathcal{G}}$. Unbarred/barred ``tangent'' indices
are raised or lowered with metrics $\mathcal{G}_{ab}$,
$\mathcal{G}_{\bar{a}\bar{b}},\ldots$, respectively. Note that in this
formulation the only infinite expansions arise from metric
inverses. Again, the equivalence of the two theories holds only up to
total derivative terms in the Lagrangian \cite{Hohm:2011dz}. However,
since we aim at describing scattering amplitudes in a flat background,
such total derivative terms are eventually irrelevant due to momentum
conservation.

In order to establish the connection between the actions
\eqref{Lsiegel} and \eqref{LDFT}, objects with ``tangent''
$GL(D)\times GL(D)$ indices have to be translated to ones with ``world
indices''. Furthermore, we are free to choose the $SO(D,D)$-frame in  
which fields do not depend on the extra coordinates
$\tilde{x}^i$. Based on \cite{Hohm:2011dz}, we thus arrive at the
following dictionary:
\begin{align}
\label{dictionary}
  h_{a\bar{b}} \rightarrow e_{ij}\, , & \qquad h^{a\bar{b}}
  \rightarrow -\frac{1}{4} e^{ij}\, ,\nonumber\\
  \langle G_{ab}\rangle  \rightarrow -2 G_{ij}\, ,\qquad 
  \langle G_{\bar{a}\bar{b}}\rangle  \rightarrow 2 G_{ij}\, , &\qquad
  \langle G^{ab} \rangle \rightarrow -\frac{1}{2} G^{ij}\, , \qquad
  \langle G^{\bar{a}\bar{b}} \rangle \rightarrow \frac{1}{2}
  G^{ij}\nonumber\\ 
  D_a  \rightarrow \partial_i\, , \qquad
  D_{\bar{a}} \rightarrow \partial_i\, , &\qquad
  D^a \rightarrow -\frac{1}{2} \partial^i\, , \qquad
  D^{\bar{a}} \rightarrow \frac{1}{2} \partial^i \, .
\end{align}
In these replacement rules the order of indices is preserved in that a
first/second index remains first/second and, hence, the manifest index
factorisation in \eqref{Lsiegel} descends directly to the equivalent
DFT Lagrangian \eqref{LDFT} (e.g.: $D^a h^{b\bar{c}} D_a h_{b\bar{c}}
\rightarrow \frac{1}{8} \partial^i e^{jk} \partial_i e_{jk}$).

\section{Perturbation theory: covariant gauge}\label{sec:cov}

The recipe for the computation of scattering amplitudes from a given Lagrangian is contained in the LSZ formalism. This instructs one to compute a time-ordered correlation function of fields, denoted $G(1,2,3,\ldots)$, amputate the external propagators and take the limit in which the momentum of the participating fields are on-shell, e.g. 
\begin{equation}
A(1,2,3,\dots) = \prod_i \lim_{p_i \rightarrow m_i^2} G(i,i)^{-1} G(1,2,3,\ldots)
\end{equation}
Behind this recipe is a careful analysis of the degrees of freedom contained in the off-shell fields of the theory. For double field theory, just as for any theory with a local gauge invariance, an ambiguity arises as there are more degrees of freedom in the off-shell fields as there are for the on-shell field content. Hence one should carefully specify the prescription for the external fields. 

To compute the needed correlation functions in the specific example under study, one first expands the DFT Lagrangian using \eqref{Lsiegel} and \eqref{dictionary} to a given order in $e_{ij}$. The kinetic terms in the
Lagrangian\footnote{For convenience, we multiply the Lagrangian from
  \cite{Hohm:2011dz} by an overall factor of $4$. Also note our convention that a
  derivative always only acts to the field immediately to its right.}
read  
\begin{equation}\label{DFT_kineticterms}
  \mathcal{L}^{(2)} = \frac{1}{4} e^{ij} {\mathcal{M}_{ij}}^{mn}
  e_{mn} +  e^{ij} \partial_i\partial_j\varphi + 
  \varphi \partial^i\partial^j e_{ij} -4\varphi \partial^i \partial_i
  \varphi \, ,
\end{equation}
where 
\begin{equation}
  {\mathcal{M}_{ij}}^{mn} = \delta_i^m
  \delta_j^n \partial^k \partial_k  - \delta_j^n \partial_i \partial^m 
  - \delta_i^m \partial_j \partial^n \, .
\end{equation}
After adding the gauge-fixing term of \cite{Hohm:2011dz} that in terms of ``world indices'' reads
\begin{equation}
  \mathcal{L}_{\text{gauge-fixing}} = \frac{1}{4} e^{jk} \left
    (\partial_k \partial^n \delta_j^m + \partial_j \partial^m
    \delta_k^n \right) e_{mn} + 2 \partial^j \varphi \partial^i e_{ij}
    -2 \partial^i \varphi \partial_i \varphi\, ,
\end{equation}
one obtains
\begin{equation}
  \mathcal{L}^{(2)} + \mathcal{L}_{\text{gauge-fixing}} = \frac{1}{4}
  e^{ij} \partial^k\partial_k e_{ij} - 2 \varphi \partial^i \partial_i
  \varphi 
\end{equation}
from which the propagators can easily be derived. Note that the kinetic term of the field $\varphi$ has the wrong sign\footnote{The
  spacetime metric has signature $(-+\ldots+)$.} which after
quantisation would lead to $\varphi$-excitations of negative energy. 

The existence of this negative energy excitation is reminiscent of the time-like component of the vector boson $A_{\mu}$ in QED (or more generally Yang-Mills theory). The unphysical part of the excitation connected to $\varphi$ must never be produced in a physical scattering process. In Yang-Mills theory this is guaranteed by gauge invariance. The story in double field theory will turn out to be more intricate, see below in subsection \ref{sub:unitarity}. 

The cubic interaction terms are (subsequently, if not stated
differently all indices are ``world indices'' despite using letters
from the beginning of the alphabet) 
\begin{align}
  \mathcal{L}^{(3)}_{e\varphi\varphi} = & -\varphi (\partial^j
  e_{ij} \partial^i \varphi + \partial^i e_{ij} \partial^j \varphi) -4
  e_{ij} \partial^i \varphi \partial^j \varphi\, ,\nonumber\\
  \mathcal{L}^{(3)}_{ee\varphi} = & \frac{1}{2} \varphi \partial_x
  e_{ab} \partial_y e_{cd} T_1^{abcdxy}\nonumber\\
  & +\frac{1}{2} e_{ij} \partial_x e_{ab} \partial_y \varphi
  T_2^{abijxy}\, ,\nonumber\\ 
  \mathcal{L}^{(3)}_{eee} = &\frac{1}{4} e_{ij} \partial_x
  e_{ab} \partial_y e_{cd} T_3^{abcdijxy}\, ,
\end{align}
where the following spacetime tensors are defined in terms of the
background Minkowski metric tensor $G_{ij}$.
\begin{align}
  T_1^{abcdxy} & = G^{ay} G^{bd} G^{cx}  - G^{ac}
  G^{bd} G^{xy} + G^{ac} G^{by} G^{dx}\nonumber\\
  T_2^{abijxy} & = G^{ay}
  G^{bj} G^{ix} + G^{ax} G^{bj} G^{iy} + G^{ai} G^{bx} G^{jy} + G^{ai}
  G^{by} G^{jx}\nonumber\\
  T_3^{abcdijxy} & = G^{ac} G^{bd} G^{ix} G^{jy} - G^{ai} G^{bd} G^{cx}G^{jy}
   - G^{ac} G^{bj} G^{dx} G^{iy}
\end{align}
Note that cubic interactions of the form $\varphi^3$ are absent.

As to quartic couplings we will furthermore explicitly give the $h^4$ ($e^4$) terms as well as the ones with $h^3 \varphi$ ($e^3 \varphi$) because
these are the ones needed for explicitly computing the tree-level amplitudes with up to five physical legs. Since they are not given in 
\cite{Hohm:2011dz} we first also give their expression in terms of ``tangent indices'': 
\begin{align}
  \mathcal{L}^{(4)}_{hhhh} = 2\Big(& -h_{c\bar{e}} D^{\bar{e}} h_{a\bar{c}}
  h^{c\bar{f}} D_{\bar{f}} h^{a\bar{c}} + h_{a\bar{e}} D^{\bar{e}}
  h_{c\bar{c}} h^{c\bar{f}} D_{\bar{f}} h^{a\bar{c}} - h_{e\bar{a}}
  D^e h_{a\bar{c}} h^{f\bar{c}} D_f h^{a\bar{a}}\nonumber \\
  &+D^c h_{a\bar{c}} D_d h^{a\bar{c}} h_{c\bar{e}} h^{d\bar{e}} -
  D_a h_{c\bar{c}} D^d h^{a\bar{c}} h^{c\bar{e}} h_{d\bar{e}} + D_c
  h_{a\bar{c}} D^c h^{a\bar{d}} h^{e\bar{c}} h_{e\bar{d}}\nonumber\\
  &- D^b h^{d\bar{c}} D_d h_{b\bar{d}} h_{e\bar{c}} h^{e\bar{d}} + D_c
  h_{a\bar{c}} D^c h^{b\bar{c}} h^{a\bar{e}} h_{b\bar{e}} - D_a
  h_{c\bar{c}} D^c h^{b\bar{c}} h^{a\bar{e}} h_{b\bar{e}} \nonumber\\
  &+D_{\bar{a}} h_{a\bar{c}} D^{\bar{c}} h^{a\bar{b}} h^{e\bar{a}}
  h_{e\bar{b}} + D_{\bar{a}} h_{a\bar{c}} D^{\bar{d}} h^{a\bar{a}}
  h^{e\bar{c}} h_{e\bar{d}} + D_{\bar{a}} h_{a\bar{c}} D^{\bar{c}}
  h^{b\bar{a}} h^{a\bar{e}} h_{b\bar{e}} \Big)
\end{align}
\begin{align}
  \mathcal{L}^{(4)}_{hhh\varphi} = &8\varphi \Big (
- h_{e\bar{a}} D^eh_{a\bar{c}} D^{\bar{c}}h^{a\bar{a}}
- h_{a\bar{e}} D^c h^{a\bar{c}} D^{\bar{e}} h_{c\bar{c}}
+ h_{c\bar{e}} D^{\bar{e}} h_{a\bar{c}} D^c h^{a\bar{c}}\Big )
\nonumber\\
&
-4D^{e}\varphi \Big (h_{a\bar{b}} h_{e\bar{c}} D^{\bar{b}}
h^{a\bar{c}}
+ h_{a\bar{c}} h^{a\bar{d}} D^{\bar{c}} h_{e\bar{d}}
+ h^{a\bar{b}} h_{e\bar{b}} D^{\bar{c}} h_{a\bar{c}}\Big)\nonumber\\
&
-4D^{\bar{e}}\varphi \Big (h_{a\bar{e}} h_{b\bar{c}} D^b h^{a\bar{c}}
+ h^{c\bar{c}} h_{c\bar{e}} D^a h_{a\bar{c}}
+ h_{a\bar{b}} h^{b\bar{b}} D^a h_{b\bar{e}}\Big)
\end{align}
In terms of ``world indices'' one finds:
\begin{equation}\label{Lagrangian4e}
  \mathcal{L}^{(4)}_{eeee} = \frac{1}{16} e_{ab} \partial_x
  e_{cd} \partial_y e_{ef} e_{gh} T_4^{abcdefghxy}\, ,
\end{equation}
and
\begin{equation}
  \mathcal{L}^{(4)}_{eee\varphi} = -\frac{1}{2}\varphi\, \partial_x
  e_{ab} \partial_y e_{cd} e_{ef} F_1^{abcdefxy}
-\frac{1}{4} \partial_y\varphi\, \partial_x e_{ab} e_{cd} e_{ef}
F_2^{abcdefxy}\, ,
\end{equation}
with the spacetime tensors
\begin{align}
  T_4^{abcdefghxy} = &  -G^{ag}G^{by}G^{ce}G^{df}G^{hx} +
  G^{ac}G^{by}G^{df}G^{eg}G^{hx}+G^{ay}G^{bd}G^{ce}G^{fh}G^{gx}
  \nonumber\\
  & -G^{ay}G^{bh}G^{ce}G^{df}G^{gx} + G^{ay}G^{bh}G^{cg}G^{df}G^{ex}
  -G^{ag}G^{bf} G^{ce}G^{dh}G^{xy} \nonumber\\
  & +G^{ag}G^{bf}G^{cy}G^{dh}G^{ex} -G^{ae}G^{bh}G^{cg}G^{df}G^{xy} +
  G^{ae}G^{bh}G^{cy}G^{df}G^{gx} \nonumber\\
  & +G^{ag}G^{bf}G^{ce}G^{dy}G^{hx} + G^{ag}G^{by}G^{ce}G^{dh}G^{fx}
  + G^{ae}G^{bh}G^{cg}G^{dy}G^{fx}\, ,
\end{align}
\begin{align}
  F_1^{abcdefxy} =\, & G^{ae} G^{bd} G^{cx} G^{fy}+G^{ac} G^{ex} \left
    (G^{by}
  G^{df}-G^{bd} G^{fy}\right ) \nonumber\\
  F_2^{abcdefxy} =\, & G^{ae} G^{by} G^{cx} G^{df}+G^{ac} G^{bx} G^{df}
  G^{ey}+G^{ay} G^{bf} G^{ce} G^{dx}\nonumber\\
  &+G^{ac} G^{bf} G^{dy}
  G^{ex}+G^{ac} G^{bf} G^{dx} G^{ey}+G^{ax} G^{bd} G^{ce} G^{fy} \, .
\end{align}
Note that despite transferring unbarred and barred ``tangent indices''
to ``world indices'', the factorisation property of the DFT Lagrangian
--- first indices of the $e_{ij}$ are never contracted into second
indices of the $e_{ij}$ --- is maintained. Furthermore, we
double-checked the quartic $e$ couplings in \eqref{Lagrangian4e} by 
directly expanding the DFT action (2.2) of \cite{Hohm:2011dz} around a
constant background and using the non-linear field redefinition (2.12)
of \cite{Hohm:2011dz}.  

As to quintic couplings we only need the $h^5 (e^5)$ interaction terms
for the explicit computation of the $e^5$ scattering amplitude. In
terms of both ``tangent indices'' and ``world indices'' one finds
\begin{align}
\mathcal{L}_{h^5}^{(5)} = 4\Big(
& h_{e\bar{a}} h^{i\bar{a}}
h_{i\bar{b}} D^e h_{a\bar{c}} D^{\bar{c}} h^{a\bar{b}} + h_{e\bar{a}}
h^{i\bar{c}} h_{i\bar{d}} D^eh_{a\bar{c}} D^{\bar{d}}h^{a\bar{a}} 
+ h_{a\bar{b}} h^{e\bar{a}} h^{b\bar{b}} D_e h^{a\bar{c}}
D_{\bar{c}} h_{b\bar{a}} \nonumber\\
& + h^{a\bar{e}} h_{a\bar{b}} h^{b\bar{b}} D_c h_{b\bar{c}}
D_{\bar{e}} h^{c\bar{c}} - h_{a\bar{b}} h^{b\bar{b}} h_{c\bar{e}}
D^{\bar{e}} h^{a\bar{c}} D^c h_{b\bar{c}} + h_{a\bar{e}}h^{c\bar{b}}
h_{d\bar{b}} D^d h^{a\bar{c}} D^{\bar{e}} h_{c\bar{c}}\nonumber\\
& - h_{c\bar{e}}
h^{c\bar{b}} h_{d\bar{b}} D^{\bar{e}} h_{a\bar{c}} D^d h^{a\bar{c}} 
+ h^{a\bar{e}} h_{i\bar{c}} h^{i\bar{d}} D_c
h_{a\bar{d}} D_{\bar{e}} h^{c\bar{c}} - h_{c\bar{e}} h^{i\bar{c}}
h_{i\bar{d}} D^{\bar{e}} h_{a\bar{c}} D^c h^{a\bar{d}} 
\Big) \, , 
\end{align}
which yields
\begin{align}
  \mathcal{L}_{e^5}^{(5)} = \frac{1}{16} e_{ab} e_{cd}
  e_{ef} \partial_xe_{gh} \partial_y e_{ij} T_5^{abcdefghijxy}\, 
\end{align}
with the spacetime tensor
\begin{align}
  T_5^{abcdefghijxy} =\, &
G^{gi} G^{ae} G^{by} G^{cx} G^{d f} G^{h j}+G^{g e} G^{i a} G^{b f} G^{c x} G^{d y} G^{h
  j}-G^{g a} G^{b y} G^{c x} G^{d f} G^{i e} G^{h j}\nonumber\\
&-G^{g a} G^{b d} G^{c e} G^{f y} G^{h j} G^{i x}-G^{g i} G^{a x} G^{j b} G^{c e} G^{h d}
G^{f y}+G^{g i} G^{a x} G^{by} G^{c e} G^{j d} G^{h f}\nonumber\\
&-G^{g i} G^{a x} G^{bf} G^{c e} G^{j d} 
G^{h y}-G^{g e} G^{i a} G^{b f} G^{c x} G^{j d} G^{h y}-G^{g a} G^{b
  y} G^{c e} G^{j d} G^{h f} 
G^{i x} \, .
\end{align}
The Mathematica notebook at \href{http://www.desy.de/~boels}{this url} can be used to further expand the Lagrangian, as required. 

\subsection{Feynman rules}
From the aforementioned terms in the Lagrangian we can read off the
following Feynman rules. The propagators for internal lines with
momentum $p$ read
\begin{eqnarray}
\label{Feynmanpropagator}
  \begin{tikzpicture}
\coordinate [] (ver1) {} ;
\coordinate [right=2cm of ver1] (ver2) {} edge [thick] node [auto,label=above : $\varphi$] {} (ver1);
\end{tikzpicture}\qquad
 &\qquad\quad & \frac{i}{4p^2}\, ,\nonumber\\
  \begin{tikzpicture}
\coordinate [label=left : $ij$] (ver1) {} ;
\coordinate [right=2cm of ver1, label=right: $mn$] (ver2) {} edge [thick] node [auto,label=above : $e$] {} (ver1);
\end{tikzpicture} & & -\frac{2i}{p^2}G_{im}G_{jn}\, .
\end{eqnarray}
Here, vertices are always defined with outgoing momenta labelled
$k_1,k_2,\ldots$ and $e$-legs are labelled with their two spacetime
indices in order to distinguish them from $\varphi$-legs. The cubic
vertices read: 
\newdimen\R
\R=1.13cm
\allowdisplaybreaks
\begin{eqnarray}
\begin{tikzpicture}
\coordinate [label=above : $ij$] () at (90:\R) {};
\draw[xshift=0\R,thick,rotate=90] (0:\R) \foreach \x in {120,240,360} {
          (\x:\R) -- (0:0)
        };
\coordinate [label=right : $k_1$] () at (90:\R/2) {};
\coordinate [label=right : $k_2$] () at (178:\R*0.85) {};
\coordinate [label=right : $k_3$] () at (345:\R*0.4) {};
\end{tikzpicture}   & & -2i k_1^i k_1^j + 4i
  (k_2^ik_3^j + k_3^i k_2^j)\, ,\nonumber\\
\begin{tikzpicture}
\coordinate [label=above : $ab$] () at (90:\R) {};
\coordinate [label=below left: $cd$] () at (210:\R) {};
\draw[xshift=0\R,thick,rotate=90] (0:\R) \foreach \x in {120,240,360} {
          (\x:\R) -- (0:0)
        };
\coordinate [label=right : $k_1$] () at (90:\R/2) {};
\coordinate [label=right : $k_2$] () at (178:\R*0.85) {};
\coordinate [label=right : $k_3$] () at (345:\R*0.4) {};
\end{tikzpicture} & & -\frac{i}{2}
  \Big (k_{1x}k_{2y}(T_1^{abcdxy} - T_2^{abcdxy} + T_1^{cdabyx} -
  T_2^{cdabyx})\nonumber\\ 
  && \qquad - k_{1x} k_{1y} T_2^{abcdxy} - k_{2x} k_{2y} T_2^{cdabxy} \Big)\,
  ,\nonumber\\
\begin{tikzpicture}
\coordinate [label=above : $ab$] () at (90:\R) {};
\coordinate [label=below left: $cd$] () at (210:\R) {};
\coordinate [label=below right: $ij$] () at (330:\R) {};
\draw[xshift=0\R,thick,rotate=90] (0:\R) \foreach \x in {120,240,360} {
          (\x:\R) -- (0:0)
        };
\coordinate [label=right : $k_1$] () at (90:\R/2) {};
\coordinate [label=right : $k_2$] () at (178:\R*0.85) {};
\coordinate [label=right : $k_3$] () at (345:\R*0.4) {};
\end{tikzpicture}   & & -\frac{i}{4} \Big (
  k_{1x}k_{2y} (T_3^{abcdijxy} + T_3^{cdabijyx} - T_3^{abijcdxy} -
  T_3^{cdijabyx} - T_3^{ijcdabxy} - T_3^{ijabcdyx})\nonumber\\
  && \qquad -k_{1x} k_{1y} (T_3^{abijcdxy} + T_3^{ijabcdxy}) - k_{2x} k_{2y}
  (T_3^{cdijabxy} + T_3^{ijcdabxy}) \Big )\, .
\end{eqnarray}
The $e^4$-vertex is
\begin{equation}
\begin{tikzpicture}
\coordinate [label=above left : $ab$] (verleg1) {} ;
\coordinate [below right=.8cm and .8cm of verleg1] (ver4) {} edge [thick] node
[auto,label=left : $k_1$] {} (verleg1);
\coordinate [below left=.8cm and .8cm of ver4,label=below left : $cd$] (verleg2) {} edge [thick] node
[auto,label=left : $k_2$] {} (ver4);
\coordinate [above right=.8cm and .8cm of ver4,label=above right : $gh$] (verleg3) {} edge [thick] node
[auto,label=right : $k_4$] {} (ver4);
\coordinate [below right=.8cm and .8cm of ver4,label=below right : $ef$] (verleg4) {} edge [thick] node
[auto,label=right : $k_3$] {} (ver4);
\end{tikzpicture}
  \begin{aligned}-\frac{i}{16}
  \Big (&k_{2x}k_{3y} \tilde{T}_4^{abcdefghxy} + k_{2x}k_{4y}
  \tilde{T}_4^{abcdghefxy} + k_{3x} k_{4y}
  \tilde{T}_4^{abefghcdxy}\nonumber\\ 
  &  + k_{1x}k_{2y} \tilde{T}_4^{efabcdghxy} +   k_{1x}k_{3y}
  \tilde{T}_4^{cdabefghxy} + k_{1x}k_{4y} \tilde{T}_4^{cdabghefxy}
  \Big)
\end{aligned}
\end{equation}
given in terms of 
\begin{equation}
  \tilde{T}_4^{abcdefghxy} = T_4^{abcdefghxy} + T_4^{ghcdefabxy} +
  T_4^{abefcdghyx} + T_4^{ghefcdabyx}\, .
\end{equation}
Vertices for the $e^3\varphi$ and $e^5$ interactions
\begin{equation}
\begin{tikzpicture}
\coordinate [label=above left : $ab$] (verleg1) {} ;
\coordinate [below right=.8cm and .8cm of verleg1] (ver4) {} edge [thick] node
[auto,label=left : $k_1$] {} (verleg1);
\coordinate [below left=.8cm and .8cm of ver4,label=below left : $cd$] (verleg2) {} edge [thick] node
[auto,label=left : $k_2$] {} (ver4);
\coordinate [above right=.8cm and .8cm of ver4] (verleg3) {} edge [thick] node
[auto,label=right : $k_4$] {} (ver4);
\coordinate [below right=.8cm and .8cm of ver4,label=below right : $ef$] (verleg4) {} edge [thick] node
[auto,label=right : $k_3$] {} (ver4);
\end{tikzpicture}
\qquad 
\newdimen\R
\R=1.13cm
\begin{tikzpicture}
\draw[xshift=0\R,thick,rotate=90] (0:\R) \foreach \x in {72,144,...,360} {
          (\x:\R) -- (0:0)
        };
\coordinate (x) at (0:0);
\coordinate [label=above : $ab$] () at (90:\R) {};
\coordinate [label=left: $cd$] () at (162:\R) {};
\coordinate [label=below: $ef$] () at (234:\R) {};
\coordinate [label=below: $gh$] () at (306:\R) {};
\coordinate [label=right: $ij$] () at (18:\R) {};
\coordinate [label=above : $k_1$] () at (110:\R/2) {};
\coordinate [label=left: $k_2$] () at (172:\R/2) {};
\coordinate [label=below: $k_3$] () at (244:\R/2) {};
\coordinate [label=below: $k_4$] () at (306:\R/2) {};
\coordinate [label=right: $k_5$] () at (0:\R/2) {};
\end{tikzpicture}
\end{equation}
are not given explicitly but they can be found in the Mathematica notebook at \href{http://www.desy.de/~boels}{this url}. They have $54$ and $1080$ terms, respectively. 

In the computation of the tree-level $e^4$-scattering amplitude it is advantageous to simplify already the cubic vertices $e^2\varphi$ and
$e^3$ by using (A) momentum conservation in order to eliminate the
dependence on what is to become an internal momentum and (B) transversality of the polarisation tensors, $\xi^{ij} k_i = \xi^{ij} k_j = 0$, at each future external leg. While these manipulations
drastically reduce the number of terms in the $e^4$-scattering amplitude, at this stage from just the DFT Lagrangian it may not be directly 
clear why transversality is expected to hold given that the gauge symmetry is of a completely different kind and, hence it is not immediate that ordinary Ward
identities should hold. On the other hand, we know that the DFT Lagrangian is physically equivalent to the low-energy theory
\eqref{lowenergyclosedstringaction} of the closed bosonic string where  the transversality must be invoked owing to conformal symmetry on the world sheet. This issue is resolved below in subsection \ref{sub:unitarity}.

Taking transversality as a given for now, one finds  
\allowdisplaybreaks
\begin{eqnarray}\label{eq:twoonsheeleoneoffshellphi}
\label{vereonsheonshvarphi}
\begin{tikzpicture}
\newdimen\R
\R=1.13cm
\coordinate [label=above : $ab^{\text{on}}$] () at (90:\R) {};
\coordinate [label=below left: $cd^{\text{on}}$] () at (210:\R) {};
\draw[xshift=0\R,thick,rotate=90] (0:\R) \foreach \x in {120,240,360} {
          (\x:\R) -- (0:0)
        };
\coordinate [label=right : $k_1$] () at (90:\R/2) {};
\coordinate [label=right : $k_2$] () at (178:\R*0.85) {};
\coordinate [label=right : $k_3$] () at (345:\R*0.4) {};
\end{tikzpicture}  
  & & ik_1\cdot k_2\,  G^{ac}G^{bd}\,
  ,\nonumber\\
\begin{tikzpicture}
\newdimen\R
\R=1.13cm
\coordinate [label=above : $ab^{\text{on}}$] () at (90:\R) {};
\coordinate [label=below left: $cd^{\text{on}}$] () at (210:\R) {};
\coordinate [label=below right: $ij$] () at (330:\R) {};
\draw[xshift=0\R,thick,rotate=90] (0:\R) \foreach \x in {120,240,360} {
          (\x:\R) -- (0:0)
        };
\coordinate [label=right : $k_1$] () at (90:\R/2) {};
\coordinate [label=right : $k_2$] () at (178:\R*0.85) {};
\coordinate [label=right : $k_3$] () at (345:\R*0.4) {};
\end{tikzpicture}  
  & & \frac{i}{4}\Big (
2 k_1^c k_1^d G^{ai} G^{bj}-k_1^d k_1^i G^{ac} G^{bj}-k_1^c k_1^j
G^{ai} G^{bd} \nonumber\\
  &&
+2 k_2^a k_2^b G^{ci} G^{dj} - k_2^b k_2^i G^{ac} G^{dj} -k_2^a k_2^j
G^{bd} G^{ci} \nonumber\\
  &&-2 k_2^b k_1^c G^{ai} G^{dj}+k_2^b k_1^i G^{ac} G^{dj}
-2 k_2^a k_1^d G^{bj} G^{ci}+k_1^d k_2^i G^{ac} G^{bj}\nonumber\\
  &&
+k_2^a k_1^j G^{bd} G^{ci} -k_2^i k_1^j G^{ac} G^{bd}+k_1^c k_2^j
G^{ai} G^{bd}-k_1^i k_2^j G^{ac} G^{bd}\Big )\, .\nonumber\\
\end{eqnarray}
Note that the $e^4$-vertex with four on-shell legs only slightly 
simplifies by means of momentum conservation. It consists of $132$
different terms. 

\subsection{Explicit amplitudes at tree level}
\subsubsection{DFT tree-level 3-point amplitude and KLT}
We have already stated that $\varphi$-scattering should not occur in order to avoid energetic instabilities. In what follows we will give
another reason: one can see already in 3-point amplitudes that setting external leg factors for $\varphi$ to zero is indeed required since
the tree-level $e^3$-amplitude is precisely the KLT 3-tensor amplitude while the ``$e\varphi^2$-amplitude'' would otherwise be non-zero. In other words, the complete tree level S-matrix of $\mathcal{N}=0$ SUGRA at three points is contained in the $e^3$ derived amplitude.

To amplify this point, the tree-level would-be scattering amplitude for $e\varphi^2$-scattering derived from direct application of the LSZ formalism is 
\begin{equation}
  \mathcal{A}_{e\varphi^2} \propto
  \xi^{ij}(k_1) \zeta(k_2) \zeta(k_3) k_{2i} k_{2j}\, ,
\end{equation}
which for arbitrary kinematical configurations vanishes only if the
external leg factors $\zeta$ associated to $\varphi$
vanish. Furthermore note that $e^2\varphi$-scattering vanishes due to
special kinematics. As to $e^3$-scattering, using momentum conservation
and transversality one finds
\begin{eqnarray}
  \mathcal{A}_{e^3} & = & \mathcal{A}_{e^3}^{abcdef}
  \xi_{ab}(k_1)\xi_{cd}(k_2)\xi_{ef}(k_3)\nonumber\\
  &\propto & \Big(-2 k_2^a k_2^b G^{ce} G^{df}+2 k_2^b k_1^c G^{ae} G^{df}- k_2^b k_1^e G^{ac} G^{df}+ k_2^b k_2^e G^{ac}
   G^{df}\nonumber\\&&\quad +2 k_2^a k_1^d G^{bf} G^{ce}-2 k_1^c k_1^d G^{ae} G^{bf}+ k_1^d k_1^e G^{ac} G^{bf}- k_1^d k_2^e
   G^{ac}G^{bf}\nonumber\\&&\quad - k_2^a k_1^f G^{bd} G^{ce}+ k_1^c k_1^f G^{ae} G^{bd}+ k_2^e k_1^f G^{ac} G^{bd}+ k_2^a
   k_2^f G^{bd} G^{ce}\nonumber\\&&\quad- k_1^c k_2^f G^{ae} G^{bd}+ k_1^e
   k_2^f G^{ac} G^{bd}\Big) \,\xi_{ab}(k_1)\xi_{cd}(k_2)\xi_{ef}(k_3)\, ,
\end{eqnarray}
which up to an overall constant equals the KLT expression,
\begin{equation}
  \left(\mathcal{A}_3^{\text{
      partial}}(1,2,3)\right)^{ace} \left(\mathcal{A}_3^{\text{
      partial}}(1,2,3)\right)^{bdf}\xi_{ab}(k_1)\xi_{cd}(k_2)\xi_{ef}(k_3)\, ,
\end{equation}
with the partial Yang-Mills theory amplitudes  defined in appendix \ref{a:partialamplitudes}. Implicitly, here we also used the KLT
relation for the polarisation tensors 
\begin{equation}
\label{KLTpolarisations}
  \xi_{ab}(k_1)=\xi_a(k_1)\xi_b(k_1)\, ,\ldots\, .
\end{equation}
An obvious consequence is that the $e^3$-amplitude satisfies the usual
Ward identities,
\begin{eqnarray}
  k_{1a}\,\mathcal{A}_{e^3}^{abcdef}
  \xi_{cd}(k_2)\xi_{ef}(k_3) & = & 0\, ,\nonumber\\
  k_{1b}\,\mathcal{A}_{e^3}^{abcdef}
  \xi_{cd}(k_2)\xi_{ef}(k_3) & = & 0\, ,
\end{eqnarray}
(cf.\ appendix \ref{a:lightcone}). In what follows we will always use $\zeta(k)=0$ for external leg factors. This naively breaks perturbative unitarity: there are graphs with $\varphi$ exchange and non-zero residues for six points and above. This issue is resolved in subsection \ref{sub:unitarity}.

\subsubsection{DFT tree-level 4-point amplitude}
The tree-level $e^4$-amplitude is given by the $e^4$-vertex as well as
$e$ and $\varphi$ exchange diagrams,
\begin{equation}
\begin{tikzpicture}
\coordinate [] (verleg1) {} ;
\coordinate [below right=.8cm and .8cm of verleg1] (ver4) {} edge [thick] node
[auto] {} (verleg1);
\coordinate [below left=.8cm and .8cm of ver4] (verleg2) {} edge [thick] node
[auto] {} (ver4);
\coordinate [above right=.8cm and .8cm of ver4] (verleg3) {} edge [thick] node
[auto] {} (ver4);
\coordinate [below right=.8cm and .8cm of ver4] (verleg4) {} edge [thick] node
[auto] {} (ver4);
\coordinate [right=2.0cm of ver4,label=left :$\displaystyle+\sum_{i=2}^4$] (bplus) {};
\coordinate [right=3.2cm of ver4] (ev1) {} ;
\coordinate [above left=.8cm and .8cm of ev1,label=left : $1$] (eleg1) {} edge [thick] node
[auto] {} (ev1);

\coordinate [below left=.8cm and .8cm of ev1,label=left : {$i$}] (eleg2) {} edge [thick] node
[auto] {} (ev1);
\coordinate [right=1cm of ev1] (ev2) {} edge [thick] node [auto,label=above : $e$] {}
(ev1);
\coordinate [above right=.8cm and .8cm of ev2] (eleg3) {} edge [thick] node
[auto] {} (ev2);
\coordinate [below right=.8cm and .8cm of ev2] (eleg4) {} edge [thick] node
[auto] {} (ev2);
\coordinate [right=2.0cm of ev2,label=left :$\displaystyle+\sum_{i=2}^4$] (bplus2) {};
\coordinate [right=3.2cm of ev2] (vv1) {} ;
\coordinate [above left=.8cm and .8cm of vv1,label=left : $1$] (vleg1) {} edge [thick] node
[auto] {} (vv1);
\coordinate [below right=.8cm and .8cm of vleg1] (vv1) {} edge [thick] node
[auto] {} (vleg1);
\coordinate [below left=.8cm and .8cm of vv1,label=left : {$i$}] (vleg2) {} edge [thick] node
[auto] {} (vv1);
\coordinate [right=1cm of vv1] (vv2) {} edge [thick] node [auto,label=above : $\varphi$] {}
(vv1);
\coordinate [above right=.8cm and .8cm of vv2] (vleg3) {} edge [thick] node
[auto] {} (vv2);
\coordinate [below right=.8cm and .8cm of vv2] (vleg4) {} edge [thick] node
[auto] {} (vv2);
\end{tikzpicture}
\end{equation}
where spacetime indices are suppressed. In our representation it
consists of $669$ terms. Using transversality of polarisations and
momentum conservation it can be written as the KLT expression in terms
of partial Yang-Mills amplitudes defined in appendix
\ref{a:partialamplitudes},   
\begin{eqnarray}
  \mathcal{A}_{e^4} &=& \mathcal{A}_{e^4}^{abcdefgh} \xi_{ab}(k_1)\xi_{cd}(k_2)\xi_{ef}(k_3)\xi_{gh}(k_4) \nonumber\\
&\propto& s \left(\mathcal{A}_4^{\text{partial}}(1,2,3,4)\right)^{aceg}
  \left(\mathcal{A}_4^{\text{partial}}(1,2,4,3)\right)^{bdhf}
  \xi_{ab}(k_1)\xi_{cd}(k_2)\xi_{ef}(k_3)\xi_{gh}(k_4)\, ,\nonumber\\
\end{eqnarray}
wher $s=(k_1+k_2)^2$ is the Mandelstam invariant. As a consequence,
the Ward identities are satisfied. In fact, in the DFT representation, 
\begin{eqnarray}
\label{amp4eward}
  k_{1a}\,\mathcal{A}_{e^4}^{abcdefgh}
  \xi_{cd}(k_2)\xi_{ef}(k_3)\xi_{gh}(k_4) & = & k_1^b\, \mathcal{A}_{e^4}^{'cdefgh}
  \xi_{cd}(k_2)\xi_{ef}(k_3)\xi_{gh}(k_4)\, ,\nonumber\\
  k_{1b}\,\mathcal{A}_{e^4}^{abcdefgh}
  \xi_{cd}(k_2)\xi_{ef}(k_3)\xi_{gh}(k_4) &=& k_1^a\, \mathcal{A}_{e^4}^{'cdefgh}
  \xi_{cd}(k_2)\xi_{ef}(k_3)\xi_{gh}(k_4)\,,
\end{eqnarray}
which due to transversality vanish upon contraction with $\xi_b(k_1)$, $\xi_a(k_1)$, respectively. Note that it is due to \eqref{vereonsheonshvarphi} that no poles arise from
$\varphi$-exchange. 

\subsubsection{DFT tree-level 5-point amplitude}
The tree-level $e^5$-amplitude has been computed from the following $81$ Feynman diagrams.

\newdimen\R
\R=1.13cm
\allowdisplaybreaks
\begin{eqnarray}
&\begin{tikzpicture}
\draw[xshift=0\R,thick,rotate=90] (0:\R) \foreach \x in {72,144,...,360} {
          (\x:\R) -- (0:0)
        };
\coordinate (x) at (0:0);
\coordinate [right=2.4cm of x,label=left :$\displaystyle+\sum_{i<j}$] (bplus) {};
\coordinate [right=0.7cm of bplus] (ev1) {} ;
\coordinate [above left=.8cm and .8cm of ev1,label=left : $i$] (eleg1) {} edge [thick] node
[auto] {} (ev1);
\coordinate [below left=.8cm and .8cm of ev1,label=left : {$j$}] (eleg2) {} edge [thick] node
[auto] {} (ev1);
\coordinate [right=1cm of ev1] (ev2) {} edge [thick] node [auto,label=above : $e$] {}
(ev1);
\coordinate [above=1.13cm of ev2] (eleg3) {} edge [thick] node
[auto] {} (ev2);
\coordinate [below=1.13cm of ev2] (eleg4) {} edge [thick] node
[auto] {} (ev2);
\coordinate [right=1.13cm of ev2] (eleg5) {} edge [thick] node
[auto] {} (ev2);
\coordinate [right=4.4cm of bplus,label=left :$\displaystyle+\sum_{i<j}$] (bplus2) {};
\coordinate [right=0.7cm of bplus2] (ev1) {} ;
\coordinate [above left=.8cm and .8cm of ev1,label=left : $i$] (eleg1) {} edge [thick] node
[auto] {} (ev1);
\coordinate [below left=.8cm and .8cm of ev1,label=left : {$j$}] (eleg2) {} edge [thick] node
[auto] {} (ev1);
\coordinate [right=1cm of ev1] (ev2) {} edge [thick] node [auto,label=above : $\varphi$] {}
(ev1);
\coordinate [above=1.13cm of ev2] (eleg3) {} edge [thick] node
[auto] {} (ev2);
\coordinate [below=1.13cm of ev2] (eleg4) {} edge [thick] node
[auto] {} (ev2);
\coordinate [right=1.13cm of ev2] (eleg5) {} edge [thick] node
[auto] {} (ev2);
\end{tikzpicture}\nonumber\\
&\begin{tikzpicture}
\coordinate [below right=2.8cm and 0.7cm of x,label=left
:$\displaystyle+\frac{1}{2}\sum_{i=1}^5\sum_{j<k;j,k\neq i}$] (bplus3) {};
\coordinate [right=0.7cm of bplus3] (ev1) {} ;
\coordinate [above left=.8cm and .8cm of ev1,label=left : $j$] (eleg1) {} edge [thick] node
[auto] {} (ev1);
\coordinate [below left=.8cm and .8cm of ev1,label=left : {$k$}] (eleg2) {} edge [thick] node
[auto] {} (ev1);
\coordinate [right=1cm of ev1] (ev2) {} edge [thick] node [auto,label=above : $e$] {}
(ev1);
\coordinate [above=1.13cm of ev2,label=above : $i$] (eleg3) {} edge [thick] node
[auto] {} (ev2);
\coordinate [right=1.13cm of ev2] (ev3) {} edge [thick] node
[auto,label=above :$e$] {} (ev2);
\coordinate [above right=.8cm and .8cm of ev3] (eleg4) {} edge [thick]
node [auto] {} (ev3);
\coordinate [below right=.8cm and .8cm of ev3] (eleg5) {} edge [thick]
node [auto] {} (ev3);
\coordinate [right=6.0cm of bplus3,label=left :$\displaystyle+\frac{1}{2}\sum_{i=1}^5\sum_{j<k;j,k\neq i}$] (bplus4) {};
\coordinate [right=0.7cm of bplus4] (ev1) {} ;
\coordinate [above left=.8cm and .8cm of ev1,label=left : $j$] (eleg1) {} edge [thick] node
[auto] {} (ev1);
\coordinate [below left=.8cm and .8cm of ev1,label=left : {$k$}] (eleg2) {} edge [thick] node
[auto] {} (ev1);
\coordinate [right=1cm of ev1] (ev2) {} edge [thick] node [auto,label=above : $\varphi$] {}
(ev1);
\coordinate [above=1.13cm of ev2,label=above : $i$] (eleg3) {} edge [thick] node
[auto] {} (ev2);
\coordinate [right=1.13cm of ev2] (ev3) {} edge [thick] node
[auto,label=above :$\varphi$] {} (ev2);
\coordinate [above right=.8cm and .8cm of ev3] (eleg4) {} edge [thick]
node [auto] {} (ev3);
\coordinate [below right=.8cm and .8cm of ev3] (eleg5) {} edge [thick]
node [auto] {} (ev3);
\end{tikzpicture}\nonumber\\
&\begin{tikzpicture}
\coordinate [below right=5.6cm and 0.7cm of x,label=left
:$\displaystyle+\sum_{i=1}^5\sum_{j<k;j,k\neq i}$] (bplus3) {};
\coordinate [right=0.7cm of bplus3] (ev1) {} ;
\coordinate [above left=.8cm and .8cm of ev1,label=left : $j$] (eleg1) {} edge [thick] node
[auto] {} (ev1);
\coordinate [below left=.8cm and .8cm of ev1,label=left : {$k$}] (eleg2) {} edge [thick] node
[auto] {} (ev1);
\coordinate [right=1cm of ev1] (ev2) {} edge [thick] node [auto,label=above : $e$] {}
(ev1);
\coordinate [above=1.13cm of ev2,label=above : $i$] (eleg3) {} edge [thick] node
[auto] {} (ev2);
\coordinate [right=1.13cm of ev2] (ev3) {} edge [thick] node
[auto,label=above :$\varphi$] {} (ev2);
\coordinate [above right=.8cm and .8cm of ev3] (eleg4) {} edge [thick]
node [auto] {} (ev3);
\coordinate [below right=.8cm and .8cm of ev3] (eleg5) {} edge [thick]
node [auto] {} (ev3);
\coordinate [right=6.0cm of bplus3,label=left :$\displaystyle+\sum_{i=1}^5\sum_{j<k;j,k\neq i}$] (bplus4) {};
\coordinate [right=0.7cm of bplus4] (ev1) {} ;
\coordinate [above left=.8cm and .8cm of ev1,label=left : $j$] (eleg1) {} edge [thick] node
[auto] {} (ev1);
\coordinate [below left=.8cm and .8cm of ev1,label=left : {$k$}] (eleg2) {} edge [thick] node
[auto] {} (ev1);
\coordinate [right=1cm of ev1] (ev2) {} edge [thick] node [auto,label=above : $\varphi$] {}
(ev1);
\coordinate [above=1.13cm of ev2,label=above : $i$] (eleg3) {} edge [thick] node
[auto] {} (ev2);
\coordinate [right=1.13cm of ev2] (ev3) {} edge [thick] node
[auto,label=above :$e$] {} (ev2);
\coordinate [above right=.8cm and .8cm of ev3] (eleg4) {} edge [thick]
node [auto] {} (ev3);
\coordinate [below right=.8cm and .8cm of ev3] (eleg5) {} edge [thick]
node [auto] {} (ev3);
\end{tikzpicture}
\end{eqnarray}

\noindent Here, $i,j,k\in\{1,2,3,4,5\}$ label the legs while their
spacetime indices are suppressed. The factors of $\frac{1}{2}$
compensate for summing identical Feynman diagrams. In our
representation the $e^5$-amplitude is given in $127.920$ terms. Note
that internal $\varphi$-lines always connect to a three-vertex with
two on-shell external $e$-legs. Hence, due to
\eqref{vereonsheonshvarphi} no poles arise from $\varphi$
exchange. This is no longer true in the case of the $e^6$-amplitude
and beyond. We numerically checked that the Ward identity is
satisfied.  

\subsubsection{On higher points}
From the above computation it is easy to see that in the perturbative computation using the double field theory, although much more efficient than the Einstein-Hilbert Lagrangian complexity still increases drastically with increasing numbers of legs. Although we have searched hard for a organising principle of these Feynman graphs into a sum over squares of Yang-Mills amplitudes as is known to exist from the KLT relations, we have been unable to find one. We interpret this to mean that the double field theory action does not make color-kinematic duality manifest - at least not in this form. It would be very interesting to find an organising principle. This also means that the motivation to push for higher order results is lacking, although our Mathematica notebook at \href{http://www.desy.de/~boels}{this url} allows one to push on using Feynman diagrams, if a motivation would be found.

\subsection{On-shell gauge invariance in double field theory and perturbative unitarity} \label{sub:unitarity}

The analysis above leaves several issues to be resolved. One is the proof of transversality: that polarisation vectors have to be orthogonal to the momentum, 
\begin{equation}
k_{i,\mu} \xi_i^{\mu}=0
\end{equation}
A second issue is the on-shell gauge invariance of the theory. Both are closely related to a third issue: the question of perturbative unitarity. 

Starting with on-shell gauge invariance, what is needed is a direct proof in DFT that replacing one polarisation vector for the $e_{ij}$ fields by its on-shell momentum makes the scattering amplitude vanish. As a warm-up it is useful to note that even in Yang-Mills theory the derivation of `transversality' involves the gauge transformation only indirectly. One considers there instead the Schwinger-Dyson equation for a single off-shell gluon field,
\begin{equation}
\langle \partial_{\mu} F^{\mu\nu} X \rangle \lfloor_{\textrm{LSZ}} = 0
\end{equation}
where $X$ stands for the other on-shell legs. Isolating the single particle pole in this expression which survives LSZ reduction and dropping the contact terms of the other fields in the correlator which do not have single particle poles in their momenta allows one to only consider the linear term,
\begin{equation}
~ \langle \left(p^2 A^{\nu} - p^{\nu} p_{\mu} A^{\mu} \right) X \rangle \lfloor_{\textrm{LSZ}} = 0 
\end{equation}
Moreover, the first term will also not contribute a single particle pole. Hence, for non-zero momentum, on-shell gauge invariance or transversality for Yang-Mills scattering amplitudes follows.

In the double field theory this derivation can be repeated, with minor modifications. The starting point is the Schwinger-Dyson equation for the $\epsilon_{ij}$ field that follows from \eqref{DFT_kineticterms}
\begin{equation}\label{eq:on-shellgaugeinv}
\langle \left(\frac{1}{2} {\mathcal{M}_{ij}}^{mn} e_{mn} +  2 \partial_i\partial_j\varphi \right) X\rangle \lfloor_{\textrm{LSZ}}  =0
\end{equation}
where all terms with multi-particle but not single particle poles in the off-shell leg have been eliminated. Now contract this field equation with $\xi^i$, the polarisation vector of the off-shell leg. Then, all terms proportional to $p_i$ will get cancelled and we obtain
\begin{equation}
\langle  \left(\xi^i p^j e_{ij} \right) X\rangle \lfloor_{\textrm{LSZ}} = 0
\end{equation}
Choosing to contract equation \eqref{eq:on-shellgaugeinv} with $\xi^{j}$ leads to the conjugate result. The upshot of this simple analysis is that scattering amplitudes computed with the double field theory action will obey on-shell Ward identities. 

Actually, there is more information in equation \eqref{eq:on-shellgaugeinv}. Contracting this with momentum $p^i$ and pushing through the derivation gives
\begin{equation}\label{eq:on-shellgaugeinvII}
\langle  \left(p^i p^j e_{ij} \right) X\rangle \lfloor_{\textrm{LSZ}} = 0
\end{equation}
Finally, consider a vector $q$ such that $q \cdot p \neq 0$. Then, contracting with $q^i q^j$ gives after dropping terms which do not survive the LSZ limit,
\begin{equation}\label{eq:on-shellgaugeinvIII}
\langle  \left(q^i p^j e_{ij} + 2 (q \cdot p) \varphi \right) X\rangle \lfloor_{\textrm{LSZ}} = 0
\end{equation}

Equations \eqref{eq:on-shellgaugeinvII} and \eqref{eq:on-shellgaugeinvIII} play a crucial role in proving unitarity of the DFT in the gauge employed in this section. Consider for this the residue at the pole of an internal propagator. This residue can, in the employed covariant gauge, be written in terms of correlators as
\begin{equation}\label{eq:residueLSZlike}
\textrm{Res} = \lim_{p^2 \rightarrow 0}  - 1/2 \langle X_L e_{ij} \rangle p^2  p^2 \langle e^{ij} X_R\rangle + 4  \langle X_L \varphi \rangle p^2  p^2 \langle \varphi X_R  \rangle 
\end{equation}
Here the $X_L$ and $X_R$ stand for the remaining parts of the correlators. Note the numeric factors are a results of the non-standard normalisations of the propagators. This is almost, but not quite, the LSZ computation for the scattering amplitudes on left and right hand side. The difference is the metric contraction, which needs to be rewritten into sums over polarisations. By completeness using polarisation vectors $\xi$ in a lightcone gauge specified by a lightcone vector $q$,
\begin{equation}
\eta_{ij} = (\sum \xi_i \xi_j ) + \frac{p_i q_j + p_j q_i}{p \cdot q} 
\end{equation}
holds, where the sum ranges over all polarisations. In Yang-Mills theory in Feynman-'t Hooft gauge, the second term drops out by on-shell gauge invariance. Using this formula, one can swap metric contractions for sums over polarisation vectors. Using equations \eqref{eq:on-shellgaugeinvII} and \eqref{eq:on-shellgaugeinvIII} it follows that equation \eqref{eq:residueLSZlike} gives
\begin{equation}\label{eq:residueLSZlikeII}
\textrm{Res} = \lim_{p^2 \rightarrow 0}  - 1/2 \langle X_L e_{ij} \rangle p^2  (\sum \xi^i \xi^k)  (\sum  \xi^j \xi^l)  p^2 \langle e_{kl} X_R\rangle 
\end{equation}
which reduces to left and right scattering amplitudes computed using only the `$e_{ij}$' fields. The minus sign of the $\varphi$ propagator is crucial to make this happen!

The analysis shows that if scattering amplitudes are computed using only the `$e_{ij}$' fields and LSZ on the outside legs, the residue of internal propagators also only involve scattering amplitudes calculated using this prescription. The $\varphi$ is therefore a ghost field, designed to kill off an unphysical degree of freedom. Novel is that this ghost field contributes already at tree level. We are not aware of a similar type tree level ghost field. The same analysis applies to residues of cut loop propagators. This saves perturbative unitarity in this gauge: only physical residues appear. 

Finally, let us study transversality: the property that the polarisation vectors must be transverse. As was shown above, in the covariant gauge the propagating degrees of freedom which appear as residues of kinematic poles are the transverse modes only. Hence, the mode $p_{\mu}\xi^{\mu}$ is not a physical degree of freedom and must be eliminated on physical grounds. More prosaically, if one leaves it in, one obtains non-gauge invariant quantities: for instance, the three point function computed in the covariant gauge would give a different result to the three point function computed in the lightcone gauge explored below. 

\subsection{String corrections in double field theory: local contributions}
A question which has been attracting some attention recently, e.g. in  \cite{Hohm:2014xsa}, \cite{Marques:2015vua} \cite{Lee:2015kba} concerns string corrections to general relativity, written in double field theory language. In this context we note that a computation which becomes straightforward once unitarity is resolved is to construct certain terms in the Lagrangian which correspond to `local' scattering amplitudes. 

Consider for instance the first correction to the closed type II superstring effective action, 
\begin{equation}\label{eq:rfour}
\sim (\alpha')^3R^4
\end{equation}
which is a certain contraction of the Riemann tensor \cite{Gross:1986iv}. These terms in perturbation theory will lead to a plethora of Feynman graphs: at order $(\alpha')^3$ these contain a single vertex from the above Lagrangian, dressed with Einstein-Hilbert-generated graphs. However for four point amplitudes, and for four points only, this particular term only contributes a \emph{local} Feynman graph: one without internal propagators. For terms of this type the connection between Lagrangian and amplitude is most direct. Similarly, a $R^n$ type term, perhaps accompanied by derivatives, will contribute locally to a $n$-point amplitude. 

Locally-contributing terms in the Lagrangian can easily be written down in the double field theory. The key observation is that in lightcone gauge, one only needs to consider terms which are a function of $e^{ij}$. Hence to reproduce the four point scattering amplitude of four gravitons sourced by the term in equation \eqref{eq:rfour} for instance, one can use the known form
\begin{equation}
\mathcal{L}_{4\textrm{ gravitons}} += T^{ijkl} T^{abcd} e_{1,ia} e_{2,jb} e_{3,kc} e_{4,ld} 
\end{equation}
where the tensors $T$ are defined for instance in \cite{Gross:1986iv}. In fact, all superstring four point scattering amplitudes are of this form. Hence, the right hand side of this expression forms the complete four point, all $e$ interactions of the double field theory. Of course, there can be more four point terms in the action involving at least one $\varphi$ field. These will however not contribute to the scattering amplitude as they do not have single-particle poles.  

In terms of the useful IIB on-shell superspace \cite{Boels:2012ie} the IIB string amplitude in a flat background can be written as 
\begin{equation}
A_{4}^{D=10}   =  \frac{\delta^{10}(K) \delta^{16}(Q)}{s \, t \, u} \left[\frac{\Gamma\left(\al s + 1\right) \Gamma\left(\al t +
1\right) \Gamma\left(\al u + 1\right) }{\Gamma\left(1- (\al s) \right) \Gamma\left(1- (\al t ) \right)
\Gamma\left(1- (\al u) \right)} \right] \, ,
\end{equation}
Here the supersymmetric delta function contains all the dependence on the polarisation vectors. Its details are unimportant; what is important here is that all $\alpha'$ corrections to scattering have a universal form. They are momentum factors times a universal Lorentz structure. Hence, 
\begin{equation}
\mathcal{L}_{4\textrm{ gravitons}} += T^{ijkl} T^{abcd}  e_{1,ia} e_{2,jb} e_{3,kc} e_{4,ld}  \sum_{i=0=j}^{\infty} a_{ij} s^i_2 s^j_3 
\end{equation}
where 
\begin{equation}
s_2 = (\alpha')^2 (s^2+t^2+ u^2)  \qquad s_3 = (\alpha')^3 (s^3+t^3+ u^3)
\end{equation}
are the two completely symmetric basis polynomials which generate all symmetric polynomials of the usual Mandelstam invariants, up to momentum conservation. Here the notation 
\begin{eqnarray}
s \left( e_{1a} e_{2b} e_{3c} e_{4d}\right) &=2  \left( \partial_{\mu} e_{1a} \partial^{\mu} e_{2b} e_{3c} e_{4d}\right) \\
t \left( e_{1a} e_{2b} e_{3c} e_{4d}\right)  &=2  \left( e_{1a} \partial_{\mu} e_{2b} \partial^{\mu}  e_{3c} e_{4d}\right)\\
u \left( e_{1a} e_{2b} e_{3c} e_{4d}\right) &=2  \left( \partial_{\mu} e_{1a}  e_{2b} \partial^{\mu} e_{3c} e_{4d}\right)
\end{eqnarray}
See \cite{Boels:2013jua} for an explanation as well the extension of this too higher points. The expansion coefficients $a_{ij}$ can be obtained to high order easily. 

It will be interesting to see what extra information can be obtained from the double field theory action by comparing to the string theory scattering amplitudes. At the very least, the lightcone gauge offers a very direct way to verify if conjectured actions satisfy basic consistency with string theory. However, the just derived expressions would only be obtained \emph{after} implementing the (correct generalisation of) the strong constraint.

\section{Perturbation theory: lightcone gauge}\label{sec:lightcone}
There are several motivations to study lightcone gauge in the double field theory. First and foremost, in lightcone gauge typically only physical degrees of freedom propagate. This usually facilitates a more direct comparison between amplitudes and actions. Lightcone gauge can also be used to study large BCFW shifts of the gravity integrand, which will lead to a proof that much of the tree level S-matrix of $\mathcal{N}=0$ supergravity is reproduced by the DFT. Interestingly, the proof holds for integrands as well. 

\subsection{Lightcone gauge propagator in DFT}
Surprisingly, we have been unable to locate an expression for the gravity lightcone gauge propagator, not even for the usual Einstein-Hilbert action. As indicated in appendix \ref{app:lightcone}, the most naive extension of the usual derivation technique of the propagator through currents fails as the resulting system of equations does not have a solution. This is a consequence of the fact that not all components of the current are needed. Below we show how to circumvent this problem in DFT by adopting a special set of coordinates. In the appendix the same analysis is applied to the Einstein-Hilbert action.

The free double field theory action, after imposing the strong constraint reads after a partial integration
\begin{equation}
\mathcal{L} = e_{i j } \left(\frac{1}{4} M^{i j k l}\right) e_{k l} + 2 \varphi \partial^{i}\partial^{ j}  e_{i j} - 4 \varphi \Box \varphi  
\end{equation}
where
\begin{equation}
M^{i jk l} = \eta^{ik} \eta^{ j l} \partial^{m} \partial_{m} - \eta^{ik} \partial^{j} \partial^{ l} -  \eta^{ j l} \partial^{i} \partial^{k}
\end{equation}
As usual, computing the propagator comes down to inverting the quadratic part of the action in one form or the other. To obtain a well-defined inverse, a gauge must be chosen. Here we choose the following variant of the lightcone gauge,
\begin{equation}\label{eq:lightconegaugechoice}
q^{i} h_{i j} = 0 = q^{ j} h_{i j}
\end{equation}
Note that in gravity often a slightly different choice is made which basically amounts to $q^{i} h_{i j}  \propto q_{ j}$ and its natural conjugate, see e.g. \cite{Goroff:1983hc}. In that paper the focus however is on deriving the lightcone Lagrangian, by integrating out non-propagating modes. Although this would be interesting to pursue in the double field theory language, here we opt to focus on the lightcone gauge propagator, for which equation \eqref{eq:lightconegaugechoice} is more convenient.

To begin, define the projector on the space orthogonal to $q$ and $p$
\begin{equation}
R_{i}{}^{ j} \equiv \eta_{i}{}^{ j} - \frac{p^{\flat}_{i} q^{ j} }{q\cdot p} - \frac{q_{i} p^{\flat, j} }{q\cdot p}
\end{equation}
where
\begin{equation}
p^{\flat}_{i} = p_{i} - \frac{p^2}{2 q p} q_{i}
\end{equation}
is a massless vector by construction. Note that
\begin{equation}
R_{i}{}^{ j}  R_{ j}{}^{k} = R_{i}{}^{k} 
\end{equation}
and
\begin{equation}
p^i R_{i}{}^{ j} = p_j R_{i}{}^{ j} = q^i R_{i}{}^{ j} = q_j R_{i}{}^{ j}  = 0
\end{equation}
The following completeness relation holds,
\begin{equation}
\eta_{i}{}^{ j}  = R_{i}{}^{ j} + \frac{p^{\flat}_{i} q^{ j} }{q\cdot p} + \frac{q_{i} p^{\flat, j} }{q\cdot p}
\end{equation}
Using this, one can write for the field $h$ in lightcone gauge,
\begin{equation}
h_{i j}  = R_{i}{}^{k} R_{ j}{}^{ l} h_{k l} +  R_{i}{}^{k} \frac{q_{ j} }{q\cdot p} h_{k l} p^{\flat, l}  +  \frac{q_{i}}{q\cdot p} R_{j}{}^{ l}  p^{\flat,k}  h_{k l}  +  \frac{q_{i} q_{j} }{(q\cdot p)^2}   p^{\flat,k} h_{k l}  p^{\flat, l}
\end{equation}
The action of $M$ on this can be computed to give
\begin{equation}
M^{ij kl} h_{kl} = (R^{i k} R^{j l} h_{k l}) p^2  -  p^{j}  R^{i, k}  h_{k l} p^{\flat, l}  - p^i R^{j l}  p^{\flat,k}  h_{k l} + (q \textrm{ - containing})
\end{equation}
where the unwritten terms with a remaining $q$ vector with either on $i$ or $j$ index will contract in the action with a field which is orthogonal to this vector in both indices. This can be inserted into the Lagrangian to obtain
\begin{equation}
\mathcal{L} = \frac{1}{4} h_{ij} \left(R^{i k} R^{j l} h_{k l} p^2  - p^{j}  R^{i, k}  h_{k l} p^{\flat, l}  - p^i R^{j l}  p^{\flat,k}  h_{k l} \right)+ 2 \varphi \, p^{\flat,k} h_{k l}  p^{\flat, l} - 4 \varphi^2 p^2
\end{equation}
For convenience, define
\begin{equation}
h_T^{ij} \equiv R^{i k} R^{j l} h_{k l}  \qquad h_{k} \equiv h_{k l} p^{\flat, l} \qquad \bar{h}_l  \equiv p^{\flat,k}  h_{k l} \qquad h \equiv p^{\flat,k} h_{k l}  p^{\flat, l}
\end{equation}
Importantly, these are independent quantum fields as long as one is careful not to contract any remaining space-time indices with either $p$ or $q$. In these coordinates, the Lagrangian reads
\begin{equation}
\mathcal{L} = \frac{1}{4} h_{T, ij} p^2 h_T^{i j}  - \frac{1}{4} h_k h^k  - \frac{1}{4}  \bar{h}_l \bar{h}^l + 2 \varphi h - 4 \varphi^2 \,  p^2
\end{equation}
where one should include the restricted metric $R_{ij}$ to perform any index contractions\footnote{In effect, this is simply the Lagrangian in terms of `lightcone' indices $(+,-, i_T)$. In this language, $R_{ij}$ is the metric in the transverse space, trivially embedded in the D-dimensional space. The advantage of the current expression is its manifest covariance.}. This form of the Lagrangian makes the computation of the propagators almost trivial, except for the correlators involving $h$ and $\varphi$. For these it is easiest to introduce sources into the Lagrangian first,
\begin{equation}
\mathcal{L}' =  2 \varphi h - 4 \varphi^2 p^2 + J h + K \varphi
\end{equation}
and compute the generating functional for the correlation functions in the standard way. For this one shifts the fields $\varphi$ and $h$ such that the linear terms in these fields vanish. This requires one to solve the equations of motion,
\begin{equation}
2 \varphi + J = 0 \qquad 2 h - 8 \varphi p^2  + K =0 
\end{equation}
which yields
\begin{equation}
 \varphi = - \frac{1}{2} J  \qquad  h  = - \frac{1}{2} K- 2 J p^2
\end{equation}
The Lagrangian then contains the following quadratic terms in the sources
\begin{equation}
\mathcal{L}' =  - \frac{1}{2} J K - J^2 p^2
\end{equation}
The correlators of the $h$ and $\varphi$ fields simply follow from this by functional differentiation. 

This leads to the following list of non-vanishing correlators,
\begin{align}
\langle h_{T,ij}, h_{T,kl} \rangle & =  - \frac{2}{p^2} R_{ik} R_{jl} \\
\langle h_{i}, h_{k} \rangle & =  2 R_{ik} \\ 
\langle \bar{h}_{j}, \bar{h}_{l} \rangle & =  2 R_{jl} \\
\langle h \varphi \rangle & =  - 1/2 \\
\langle h h \rangle & =  - 2 p^2 
\end{align}
These results can be translated back into the original variables, yielding
\begin{align}
\langle h_{ij} \varphi \rangle &= - \frac{1}{2} \frac{q_{i} q_{j} }{(q\cdot p)^2}  \\
\langle h_{ij}, h_{kl} \rangle & =  - \frac{2}{p^2} R_{ik} R_{jl} + 2 R_{i k} \frac{q_{j} q_{l} }{(q\cdot p )^2} + 2  \frac{q_{i} q_{k} }{(q\cdot p )^2} R_{j l}  - 2 p^2  \frac{q_{i} q_{k} q_{j} q_{l} }{(q\cdot p )^4}
\end{align}
By re-expressing the projector $R$ in terms of $p$ and $q$ the latter correlator can also be written as
\begin{equation}\label{eq:lightconepropexp}
\langle h_{ij}, h_{kl} \rangle  =  - \ii \frac{2}{p^2} \left(\eta_{ik} - \frac{q_i p_k + q_k p_i }{q \cdot p} \right) \left(\eta_{jl} - \frac{q_j p_l + q_l p_j }{q \cdot p} \right)
\end{equation}
whose numerator is simple the square of the numerator of the usual Yang-Mills lightcone gauge propagator. Hence, in lightcone gauge, the double field theory two point function obeys a direct form of color-kinematic duality. Note that the lightcone gauge propagator in Einstein-Hilbert gravity, derived in appendix \ref{app:lightcone}, does not have this property. 

Note the absence of a $\langle \varphi  \varphi \rangle$ correlator or a pole in the $\langle \varphi h \rangle$ correlator: in lightcone gauge the $\varphi$ field does not propagate and it is purely an auxiliary field providing contact interactions in perturbation theory in this gauge. Hence the only source of single-particle poles is in the $\langle h_{T,ij}, h_{T,kl} \rangle$ correlator. Therefore these fields carry all the physical degrees of freedom. Since these are effectively $D-2 \times D-2$ tensors, this is consistent with the generic expectation that in lightcone gauge only physical degrees of freedom propagate. Stronger, in the absence of un-physical singularities associated with the factors of $\frac{1}{q\cdot p}$ in loops, the perturbation theory is manifestly unitary to all loop orders. Potential singularities from the $\frac{1}{q\cdot p}$ factors would be regulated in practice using the Mandelstam-Leibbrandt prescription, see \cite{Leibbrandt:1987qv}.

In practice, when computing a scattering amplitude in this gauge, one should insert $h_{ij}$ fields on the external legs as these are the only fields to yield the single-particle poles needed for a non-trivial LSZ reduction. Hence in this gauge one can reduce the residues at poles directly to strict $h$ correlators, in contrast to the Feynman-'t Hooft like gauge explored above where the Schwinger-Dyson equations were needed. We have verified explicitly that with the above lightcone gauge propagators the four point amplitude is correctly obtained. In fact, the computation even works without the lightcone condition on $q$ which would correspond to an axial-type gauge.

\subsection{BCFW on-shell recursion: general setup}
As in \cite{ArkaniHamed:2008yf}, we set up the BCFW on-shell recursion
\cite{Britto:2004ap,Britto:2005fq} for scattering amplitudes in $(D\ge
4)$ spacetime dimensions as follows: We analytically continue the
momenta of leg $1$ and $2$ such that they are kept on-shell and momentum 
conservation is maintained, i.e.\ the BCFW shift is taken to be 
\begin{equation}
\label{BCFW:shift}
  k_1 \rightarrow \hat{k}_1=k_1+q z\,, \qquad k_2 \rightarrow \hat{k}_2=k_2 - q z\, ,
\end{equation}
for a $q$ with $q\cdot k_{1,2}=0$ and $q^2=0$ and a complex parameter
$z$. Without loss of generality the $D$-dimensional momenta of leg 1
and 2 can be chosen to be back-to-back,
\begin{equation}
  k_1 = (1,1,0,0,\vec{0}), \qquad k_2 = (1,-1,0,0,\vec{0})\, 
\end{equation}
and $q = (0,0,1,i,\vec{0})$ must necessarily be complex. In order for
the $e$-polarisations $\xi$ to remain transversal to their momenta,
they need to be analytically continued too. For legs $1$ and $2$ the
shifts for the polarisations are given by \eqref{KLTpolarisations} and
\begin{align}
\label{BCFW:shiftpolarisations}
  \xi^-(k_1) = \xi^+(k_2)=q &\rightarrow  \hat{\xi}^-(\hat{k}_1)  = \hat{\xi}^+(\hat{k}_2)=q\, , \nonumber\\
  \xi^+(k_1)=q^* &\rightarrow \hat{\xi}^+(\hat{k}_1)  = q^* -zk_2\, , \nonumber\\
  \xi^-(k_2)=q^* &\rightarrow \hat{\xi}^-(\hat{k}_2)  = q^* +zk_1\, , \nonumber\\
  \xi^T(k_1) = \xi^T(k_2) &= \hat{\xi}^T(\hat{k}_1)
  =\hat{\xi}^T(\hat{k}_2)=(0,0,0,0,\ldots,1,\ldots,0)\, , 
\end{align}
with $SO(D-2)$ little group indices $\{-,+,T\}$. Using on-shell gauge invariance, the $z$-independent polarisation vectors result in $\frac{1}{z}$ suppressed contributions. Given an amplitude
$\mathcal{A}$, let $\hat{\mathcal{A}}$ denote the analytically
continued amplitude obtained after the shifts \eqref{BCFW:shift} and 
\eqref{BCFW:shiftpolarisations}. Then the BCFW formula
reads\footnote{Here, $\oint_{z=0}'=\frac{1}{2\pi i} \oint_{z=0}$  
  and the contour around $z=0$ is anticlockwise.}
\begin{align}
\label{BCFWformula}
  \mathcal{A} = &\oint_{z=0}' \frac{\hat{\mathcal{A}}}{z}\, dz =
  -\sum_{I} \text{Res}_{z=z_I} \frac{\hat{A}}{z} +
  \text{Res}_{z\rightarrow \infty}
  \frac{\hat{\mathcal{A}}}{z} \, ,
\end{align}
where one sums over all subsets $I\subset \{k_2,\ldots,k_n\}$  of the
set of external momenta (excluding $k_1$) such that
\begin{equation}
  z_I=\frac{\left (\sum_{i\in I} k_i \right)^2}{2 \sum_{i\in I}
    k_i\cdot q}\in\mathbb{C}\backslash\{0\}
\end{equation}
is non-zero.
The BCFW recursion for an $e^n$-amplitude $\mathcal{A}$ is possible if
the residue at infinity in \eqref{BCFWformula} vanishes. As derived in this form first in \cite{ArkaniHamed:2008yf}, the large BCFW shift behaviour, $z \rightarrow \infty$, gives
\begin{equation}\label{eq:knowngravityshift}
\lim_{z \rightarrow \infty} A(z) \propto  z^2 \hat{\xi}_{1, mn} \hat{\xi}_{2, kl}  \left(\eta^{mk} + \frac{1}{z} B^{mk} + \mathcal{O}\left( \frac{1}{z^2} \right)  \right) \left(\eta^{nl} + \frac{1}{z} B^{nl} + \mathcal{O}\left( \frac{1}{z^2} \right)  \right)
\end{equation}
where the tensor $B$ is antisymmetric in its indices. This takes the form of the square of the behaviour of two Yang-Mills amplitudes and guarantees that for any helicity on legs $1$ and $2$ a BCFW shift exists for which $A(z) \sim \frac{1}{z^2}$. This proves the existence of BCFW on-shell recursions relations. If this behaviour can be proven for the DFT Lagrangian, then the S-matrix must coincide with that of Einstein Hilbert gravity by unitarity since the three point amplitude agree. Before investigating this in more detail let us first discuss  BCFW on-shell recursion for the $e^4$ and $e^5$-amplitude.

One explicitly finds that the amplitude
\begin{equation}
\label{BCFWAe4--++}
  \mathcal{A}_{e^4}^{--++} = \mathcal{A}_{e^4}^{abcdefgh}\,
  \xi_{ab}^{--}(k_1) \xi_{cd}^{++}(k_2)\xi_{ef}(k_3)\xi_{gh}(k_4)
\end{equation}
scales with $1/z^2$ as $z\rightarrow \infty$ and thus the residue at 
infinity in \eqref{BCFWformula} vanishes. Hence --- in the absence of 
$\varphi$-poles --- $\mathcal{A}_{e^4}^{--++}$ can be expressed in terms
of analytically continued $e^3$-amplitudes:
\begin{align}
\label{BCFWrecursione4}
  \mathcal{A}_{e^4}^{--++} = &\:
  \hat{\mathcal{A}}_{e^3}^{abefij}\hat{\xi}_{ab}^{--}(\hat{k}_1)
  \xi_{ef}(k_3)  \frac{-2i G_{im}G_{jn}}{t}
  \hat{\mathcal{A}}_{e^3}^{cdghmn}\hat{\xi}_{cd}^{++}(\hat{k}_2)\xi_{gh}(k_4)\Big
  |_{z=\frac{t}{2k_4\cdot 
    q}}\nonumber\\
  & +
  \hat{\mathcal{A}}^{abghij}_{e^3}\hat{\xi}_{ab}^{--}(\hat{k}_1)\xi_{gh}(k_4)
  \frac{-2iG_{im}G_{jn}}{u} \hat{\mathcal{A}}^{cdefmn}_{e^3}
  \hat{\xi}_{cd}^{++}(\hat{k}_2) \xi_{ef}(k_3) \Big |_{z=\frac{u}{2k_3\cdot
    q}}\, .
\end{align}
Pictorially, this is
\begin{equation}
\begin{tikzpicture}
\coordinate [right=2.0cm of ver4,label=left
:${\displaystyle\mathcal{A}_{e^4}^{--++} = \sum_{i=3}^4}$] (bplus) {}; 
\coordinate [right=3.2cm of ver4] (ev1) {} ;
\coordinate [above left=.8cm and .8cm of ev1,label=above : $\hat{1}^{--}$] (eleg1) {} edge [thick] node
[auto] {} (ev1);
\coordinate [below left=.8cm and .8cm of ev1,label=below left : {$i$}] (eleg2) {} edge [thick] node
[auto] {} (ev1);
\coordinate [right=1.3cm of ev1] (ev2) {} edge [thick] node [auto,label=above : $e$] {}
(ev1);
\coordinate [above right=.8cm and .8cm of ev2,label=above right:$\hat{2}^{++}$] (eleg3) {} edge [thick] node
[auto] {} (ev2);
\coordinate [below right=.8cm and .8cm of ev2] (eleg4) {} edge [thick] node
[auto] {} (ev2);
\fill (ev1) circle (0.25);
\fill (ev2) circle (0.25);
\end{tikzpicture}\Big|_{z=-\frac{s_{1i}}{2k_i\cdot q}}\, .
\end{equation}
Note that in \eqref{BCFWrecursione4} the $e$-propagator as in 
\eqref{Feynmanpropagator} appears. Other helicity configurations can
be checked analogously. 

We  also checked numerically that the large-$z$ scaling for the
$\mathcal{A}_{e^5}^{--++}$ amplitude defined analogously as in
\eqref{BCFWAe4--++} is again $1/z^2$. Hence this amplitude can be computed by summing over the residues of the following 6 channels, 
\begin{equation}
\label{BCFWrecursione5}
\begin{tikzpicture}
\coordinate [right=2.0cm of ver4,label=left
:${\displaystyle\mathcal{A}_{e^5}^{--++} = \sum_{i=3}^5}$] (bplus) {}; 
\coordinate [right=3.2cm of ver4] (ev1) {} ;
\coordinate [above left=.8cm and .8cm of ev1,label=above : $\hat{1}^{--}$] (eleg1) {} edge [thick] node
[auto] {} (ev1);
\coordinate [below left=.8cm and .8cm of ev1,label=below left : {$i$}] (eleg2) {} edge [thick] node
[auto] {} (ev1);
\coordinate [right=1.3cm of ev1] (ev2) {} edge [thick] node [auto,label=above : $e$] {}
(ev1);
\coordinate [above=.8cm of ev2,label=above:$\hat{2}^{++}$] (eleg3) {} edge [thick] node
[auto] {} (ev2);
\coordinate [right=.8cm and .8cm of ev2] (eleg4) {} edge [thick] node
[auto,label=below right : $\quad\Big|_{z=-\frac{s_{1i}}{2k_i\cdot q}}$] {} (ev2);
\coordinate [below=.8cm and .8cm of ev2] (eleg4) {} edge [thick] node
[auto] {} (ev2);
\fill (ev1) circle (0.25);
\fill (ev2) circle (0.25);
\coordinate [right=3.5cm of ev2,label=left
:${\displaystyle+\sum_{i=3}^5}$] (bplus) {}; 
\coordinate [right=4.7cm of ev2] (ev1) {} ;
\coordinate [above left=.8cm and .8cm of ev1,label=above : $\hat{2}^{++}$] (eleg1) {} edge [thick] node
[auto] {} (ev1);
\coordinate [below left=.8cm and .8cm of ev1,label=below left : {$i$}] (eleg2) {} edge [thick] node
[auto] {} (ev1);
\coordinate [right=1.3cm of ev1] (ev2) {} edge [thick] node [auto,label=above : $e$] {}
(ev1);
\coordinate [above=.8cm of ev2,label=above:$\hat{1}^{--}$] (eleg3) {} edge [thick] node
[auto] {} (ev2);
\coordinate [right=.8cm and .8cm of ev2] (eleg4) {} edge [thick] node
[auto,label=below right : $\quad\Big|_{z=\frac{s_{2i}}{2k_i\cdot q}}$] {} (ev2);
\coordinate [below=.8cm and .8cm of ev2] (eleg4) {} edge [thick] node
[auto] {} (ev2);
\fill (ev1) circle (0.25);
\fill (ev2) circle (0.25);
\end{tikzpicture}\, .
\end{equation}

\subsection{BCFW on-shell recursion: large BCFW shifts in DFT}
To verify the S-matrix generated by the double field theory, one can try to verify that the large BCFW shift behaviour of the tree level scattering amplitude reproduces the known result \eqref{eq:knowngravityshift}. More generally, one can try to prove a similar result for the loop level integrand, in order to verify a conjecture in \cite{Boels:2010nw} that the integrand can, in principle, be reconstructed from its single cuts. The technique to be used below is the same as applied to Yang-Mills theory in \cite{Boels:2010nw}: consider the Feynman diagrams in the natural lightcone gauge specified by the BCFW shift vector. The usefulness of this gauge choice was first advocated in \cite{ArkaniHamed:2008yf}. 

The Feynman diagrams split into two categories: those where the two shifted legs appear on a three vertex and those where this does not happen. The reason the three-vertex case is special is that the momentum flowing through the off-shell leg of these Feynman diagrams is orthogonal to the shift vector: $q \cdot (k_1 + k_2) = 0$ and therefore the lightcone gauge choice $q$ is at face value illegal for this class of Feynman graphs. As explained in \cite{Boels:2010nw} for Yang-Mills theory, this class of diagrams can be treated by first imposing a lightcone gauge choice using the vector $\tilde{q} = q+ x k_1$, and then sending $x$ to zero. 

For the case at hand, there are three Feynman graphs with the shifted legs connected to the three vertex: one with a $e^2$ propagator and two with the mixed propagator $e\varphi$. Since 
\begin{equation}
\tilde{q} \cdot (k_1 + k_2) = x (k_1 \cdot k_2)
\end{equation}
both propagators have up to a second order pole in $x$. This is however a fictitious pole: gauge invariance of the overall expression ensures that the full result is independent of the gauge choice used to compute it.  First consider the graph with the $\varphi$ field off-shell. The vertex with two on-shell $e$ fields is listed in equation \eqref{eq:twoonsheeleoneoffshellphi}; it's structure, ``$\propto z^0 \eta \eta$'', confirms to equation \eqref{eq:knowngravityshift}. Taken together with the $\varphi e$ propagator this contribution is proportional to
\begin{equation}\label{eq:evarphilargeshifts}
\propto  \eta_{ac} \eta_{bd} \frac{q^{k}q^{l}}{x^2 (k_1 \cdot k_2)}
\end{equation}
where the indices $k,l$ are contracted into the vertex the propagator connect to (this is not the three vertex which the shifted legs connect to). 

The vertex with two on-shell and one off-shell $e$ field is special. Since in the on-shell limit this reproduces the three point graviton scattering amplitude, it's structure is very closely related to the `square' of two Yang-Mills vertices. In fact, the difference of the product of two three point Yang-Mills vertices with one off-shell leg and that the DFT three point vertex with one off-shell leg is:
\begin{equation}
(DFT)_3 - ((YM)_3)^2 = - 4 (k_{1}+k_2)^i  (k_{1}+k_2)^j \eta_{ac} \eta_{bd}
\end{equation}
Here $a,b$ and $c,d$ are the indices of the shifted legs and $i,j$ are the indices of the off-shell leg. When contracted into the $e\varphi$ propagator, this is a pure, $x$-independent contact terms whose structure is
 ``$\propto z^0 \eta \eta$''.  When contracted into the $ee$ propagator, the propagator almost vanishes, leaving
\begin{equation}
\propto -4 \eta_{ac} \eta_{bd} \frac{q^{k}q^{l}}{ 4 x^2 ( k_1 \cdot k_2)}
\end{equation}
The remaining terms are simply the square of the Yang-Mills result in the large $z$ limit, written out explicitly in equation 2.12 in \cite{Boels:2010nw}. This follows since the vertex is basically such a square and the $ee$ lightcone gauge propagator in DFT is concerning its numerator the square of the Yang-Mills result.  In all, this shows that the three point contribution to the large z limit in DFT is exactly captured by the known gravity result, equation \eqref{eq:knowngravityshift}.

Let us  study further classes of diagrams contributing to the large $z$ behaviour. There are two sources of positive powers of $z$ which are always accompanied by the gauge choice vector $q$. These two sources are the lightcone gauge propagator as well as the vertices themselves. Both come maximally with one power of $z$, but many graphs contribute at this order.  Note though that both of these two sources of positive powers of $z$ are, in a real sense, orthogonal: the positive power of $z$ in the propagator comes with four $q$ vectors which contract into the vertices. Since this vector should not contract into another $e$ field\footnote{This vector can contract into the shifted polarisation vectors, for which a single negative power of $z$ is obtained.}, for maximal contribution these should contract into the momenta of the vertex. Hence, the two positive sources of $z$'s are orthogonal. 

\subsubsection{DFT S-matrix from on-shell recursion}
Note that this is enough already to prove BCFW on-shell recursion at tree level for a large class of scattering amplitudes with external gravitons. In $D$ dimensions, two particle on-shell momenta and a choice of BCFW shift vector span a three dimensional space. Given the polarisation vector of the graviton, one can always choose a BCFW shift vector to it into a special set of four dimensions. Hence without loss of generality we can assume one of the graviton legs is in four dimensions. There is always a BCFW shift for which the polarisation vector of a graviton scales as $\frac{1}{z^2}$. As long as the other BCFW-shifted leg does not have the exact same graviton polarisation as the other leg, its scaling behaviour is $z^0$ or better.  Taken together with the result that the leading, $z^2$, part of the large BCFW shift scales with two metrics, this means that for the other shifted legs any graviton polarisation one awayls obtains a $\frac{1}{z}$ fall-off of the scattering amplitude. 

All scattering amplitudes with only gravitons have at least one opposite pair of helicities. The crucial observation is that if all external particles have the same helicities, then one can choose a gauge for which like helicity polarisation vectors are orthogonal,
\begin{equation}
\xi^{\pm}_{1,\mu} \xi^{\pm, \mu}_2 = 0 
\end{equation}
holds. It is easy to check that in every Feynman graph, there must be at least half of the indices of external particles being contracted. Hence, like helicity amplitudes (all-plus) vanish. As check above the three point amplitude matches between DFT and Einstein gravity. Therefore using BCFW on-shell recursion shows the complete tree-level S-matrix of Einstein-Hilbert gravity is reproduced from the double field theory.  

This result easily extends to the integrand of gravity amplitudes in supersymmetric theories - for which it is known there are no like-helicity scattering amplitudes. However, this only shows that the integrand may be reconstructed from lower-loop amplitudes as well as single cuts. See \cite{Boels:2010nw} for a discussion.

\subsubsection{Explicit off-shell sub-cancellations using index ordering}

\subsubsection*{Vertex dependence}
In order to look for further structure and to compare to Yang-Mills theory through the double copy construction, we have studied sub-leading terms in the large $z$ limit. To spot structure in the graphs, it pays to pay attention to index contractions within the graphs. For instance, within the vertices a positive power of $z$ can only arise if a $z$ dependent hard momentum (i.e. either $p_1$ or $p_2$) is contracted directly into a momentum of one of the other off-shell legs; otherwise, a $q$ contracts either into a leg orthogonal to $q$ due to the lightcone propagator, or it contracts directly into a shift vector (which gives a $z$-suppressed contribution). This leaves an index contraction between all the legs, which looks remarkably like a color-ordered amplitude. Note that the number of legs must be even, and the vertex contributing at this order reads
\begin{equation}\label{eq:largezverts}
\lim_{z\rightarrow \infty} V_n =  (1+\varphi)^2 \left(\partial^{m} e_{ab} \partial_{m} e_{cd} G^{ac} G^{db}  \right) + \mathcal{O}\left(z^0 \right)
\end{equation}
First consider the contributions without $\varphi$ field. Since the propagator connecting vertices of this type will be $\frac{1}{z}$, there are many contributions for a given number of off-shell external legs connecting to the hard line graph. These can be \emph{ordered} and labelled consecutively. The intuition is that it is hard for two contributions with interchanged legs to cancel in any meaningful way. This is important as it gives a group-theory-like structure to the large $z$ limit graphs. We will refer to this as "index-ordering". 

Consider now four point graph contributions generated by the above vertex: two of the legs are BCFW shifted. However, we will assume the BCFW shifted legs appear on the right or left side of the graphs, as a part of a larger current. The momenta of these currents will be labelled $a$ and $b$ respectively. Note $q\cdot p_a$ is only zero if this current contains only one leg. There are two graph topologies: either expanding the `left' or the `right' vertex in equation \eqref{eq:largezverts}. The results are
\begin{equation}
\lim_{z\rightarrow \infty}  G_4(a,1,2,b) = -2 z (p_2 \cdot q + p_a \cdot q) \eta_R (e^2)_L + \mathcal{O}\left(z^0 \right)
\end{equation}
and
\begin{equation}
\lim_{z\rightarrow \infty} G_4(a,b,1,2) = -2 z (p_1 \cdot q + p_a \cdot q) \eta_L (e^2)_R + \mathcal{O}\left(z^0 \right)
\end{equation}
where $G$ is the four point Green's function whose arguments are index-ordered. Let us stress that all the legs are off-shell. For all legs on-shell it is easy to cross-check that the above contribution is anti-symmetric under interchange of the indices not associated to the metric contraction: this confirms the earlier statement about the shift of the four point amplitude.

For six point graph contributions, there are three index-ordered contributions possible. There are two which involve a metric contraction between the $a$ and $b$ legs. These read
\begin{equation}
\lim_{z\rightarrow \infty}  G_4(a,1,2,3,4,b) = -2 z \frac{(p_1 \cdot q )(p_4 \cdot q )}{q\cdot(p_1+p_2+p_a)} \eta_R (e^4)_L + \mathcal{O}\left(z^0 \right)
\end{equation}
and for the conjugate
\begin{equation}
\lim_{z\rightarrow \infty}  G_6(a,b, 4,3,2,1) = -2 z \frac{(p_1 \cdot q)(p_4 \cdot q)}{q\cdot(p_3+p_4+p_a)} \eta_L (e^4)_R + \mathcal{O}\left(z^0 \right)
\end{equation}
Note that at order $z^0$, the index structure of the non-metric term is irrelevant, see equation \eqref{eq:knowngravityshift}.

\begin{figure}[!htb]
\centering 
\includegraphics[scale=0.5]{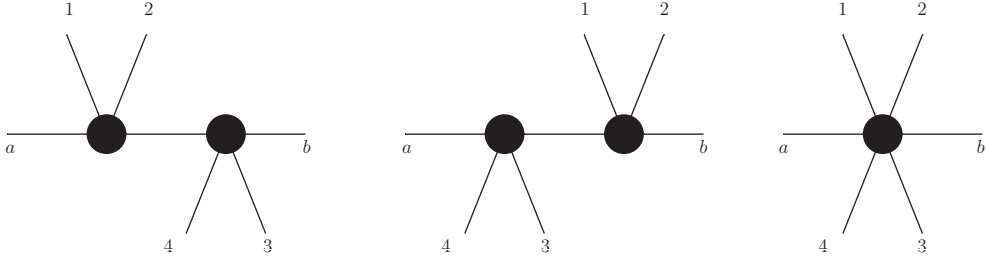}
\caption{\label{fig:ii} Six point index-ordered graph contributions with two sets of legs on opposite sides .}
\end{figure}

The remaining contraction is $G_4(a,1,2,b,3,4)$, which involves summing the three Feynman graphs in figure \ref{fig:ii}. Taking $a$ to be a BCFW shifted leg so that $p_a \cdot q = 0$,  we obtain
\begin{equation}
\lim_{z\rightarrow \infty}  G_6(a,1,2,b,3,4) \propto  z \frac{(p_2 \cdot q) (p_3 \cdot q) (q\cdot (p_1+p_2+p_3+p_4))}{q\cdot(p_3+p_4) \,\,q\cdot(p_1+p_2)} +  \mathcal{O}\left(z^0 \right)
\end{equation}
Hence, if leg $b$ would be taken to be the other BCFW shifted leg, the order $z$ contribution of these graphs vanishes. The legs $1$ through $4$ remain off-shell: this shows cancellations in a sub-class of Feynman graphs. Taking $a$ and $b$ to be BCFW shifted legs, it is interesting to check the index structure of the order $z^0$ term: according to equation \eqref{eq:knowngravityshift} this should either involve a metric contraction on the left or right sides between the indices of the shifted legs, or it should be antisymmetric under inversion of the indices of either the left or right sides. Since there is no metric contraction in $G_6(a,1,2,b,3,4)$, one should study the symmetry properties of the index-ordered expressions. The symmetry in indices comes down to checking inverting the order on one of the sides of the index ordered expression. One computes
\begin{equation}
\lim_{z\rightarrow \infty}  G_6(a,1,2,b,3,4) + \lim_{z\rightarrow \infty}  G_6(a,2,1,b,3,4) = 0  +  \mathcal{O}\left(z^{-1} \right)
\end{equation}
which shows that as expected the order $z^0$ index contractions are anti-symmetric in the left side. The computation involving an inversion of $34$ follows analogously. We have verified the sub-leading, $ \mathcal{O}\left(z^{-1} \right)$, term in this computation is indeed antisymmetric in momenta $3$ and $4$ or in momenta $1$ and $2$ (or in both), needed for the  $\mathcal{O}\left(z^{-1} \right)$ term in equation \eqref{eq:knowngravityshift} . 

For eight point graph contributions, there are four index-ordered contributions possible, two of which are related by left-right symmetry. Taking $p_a \cdot q = 0$, 
\begin{multline}
\lim_{z\rightarrow \infty} G_8(a,1,2,3,4,5,6,b) = \\-2 z \frac{(p_1 \cdot q  )(p_6 \cdot q)(q\cdot(p_1+p_2+p_3))}{(q\cdot(p_1+p_2))(q\cdot(p_1+p_2+p_3+p_4))} \eta_R (e^6)_L + \mathcal{O}\left(z^0 \right)
\end{multline}
and
\begin{multline}
\lim_{z\rightarrow \infty}  G_8(a,1,2,3,4,b,5,6)  \propto  \\ z \frac{(p_1 \cdot q) (p_4 \cdot q)  (p_5 \cdot q) (q\cdot (p_1+p_2+p_3+p_4+p_5+p_6))}{q\cdot(p_1+p_2) \,\,q\cdot(p_1+p_2+ p_3+p_4) \,\, q\cdot(p_5+p_6) }  \mathcal{O}\left(z^0 \right)
\end{multline}
are obtained. If leg $b$ is also a BCFW-shifted leg, the latter expression vanishes. This is the result of summing $8$ Feynman graphs. Furthermore,
\begin{equation}
\lim_{z\rightarrow \infty}  G_8(a,1,2,3,4,b,5,6) + \lim_{z\rightarrow \infty}  G_8(a,1,2,3,4,b,6,5)  = 0  +  \mathcal{O}\left(z^{-1} \right)
\end{equation}
and 
\begin{equation}
\lim_{z\rightarrow \infty}  G_8(a,1,2,3,4,b,5,6) + \lim_{z\rightarrow \infty}  G_8(a,4,3,2,1,b,5,6)  = 0  +  \mathcal{O}\left(z^{-1} \right)
\end{equation}
hold. We have verified the sub-leading, $ \mathcal{O}\left(z^{-1} \right)$, term in both computations is indeed antisymmetric in reversing $5$ and $6$ or in $(1,2,3,4)$ (or both), needed for the  $\mathcal{O}\left(z^{-1} \right)$ term in equation \eqref{eq:knowngravityshift} . 

For ten or more graphs the combinatorics gets more and more complicated. One class of diagrams which is easier are those with one metric contraction. In fact, we strongly suspect that
\begin{multline}\label{eq:leadinglargezwsinglemetric}
\lim_{z\rightarrow \infty}  G_{n+2} (a,1,\ldots,n,b) = \\-2 z \frac{(p_n \cdot q)(q \cdot p_1 )(q\cdot(p_1+p_2 + p_3 + p_a)\ldots (q\cdot(p_1+\ldots + p_{n-1}+ p_a))}{(q\cdot(p_1+p_2 + p_a))(q\cdot(p_1+p_2+p_3+p_4 + p_a)) \ldots (q\cdot(p_1+\ldots+p_{n-2}+ p_a))} \\ \eta_R (e^n)_L + \mathcal{O}\left(z^0 \right)
\end{multline}
holds, together with its natural conjugate. This we checked through $n=14$, where the last involves summing $64$ Feynman graphs. Furthermore, we obtained
\begin{multline}
\lim_{z\rightarrow \infty} G_8(a,1,\ldots,6,b,7,8) = \lim_{z\rightarrow \infty}  G_8(a,1\ldots,4,b,5,\ldots, 8)  = \\
 \lim_{z\rightarrow \infty}  G_{10}(a,1,\ldots,8,b,9,10) = \lim_{z\rightarrow \infty} = G_{10}(a,1,\ldots,6,b,7,\ldots,10) =  \mathcal{O}\left(z^0 \right) 
\end{multline}
for BCFW-shifted legs $a$ and $b$.

Adding $\varphi$ legs does not alter any of statements above on the order of $z$ contributions or on the index structure. To see this derive the Schwinger-Dyson equation for the $e$ field, dropping all contact terms as before,
\begin{equation}
\langle \frac{1}{2} M^{ijkl} e_{kl} + 2 p^i p^j \varphi + \frac{\delta}{\delta e_{ij}} \mathcal{L}_{n>2} | X \rangle = 0 
\end{equation}
where $X$ stands for an arbitrary collection of on-shell fields and we have split the contributions into those of the quadratic Lagrangian and those beyond, with the latter denoted $ \mathcal{L}_{n>2} $. The last term is effective singling out one particular leg of a vertex in a Feynman graph computation. In lightcone gauge, the first two terms lead to very particular further contributions which can be written in similar notation. First contract the equation with a single momentum,
\begin{equation}
\langle - \frac{1}{2} p^j (p^k p^l e_{kl}) + 2 p^2 p^j \varphi  + p^{i} \frac{\delta}{\delta e_{ij}} \mathcal{L}_{n>2} | X \rangle = 0 
\end{equation}
Then, use the explicit form of the lightcone gauge propagators to derive
\begin{equation}\label{eq:schwingdysimp}
\langle  p^j \frac{\delta}{\delta \varphi} \mathcal{L}_{n>2}+ p^{i} \frac{\delta}{\delta e_{ij}} \mathcal{L}_{n>2} | X \rangle = 0 
\end{equation}
Again, the zero on the RHS is the absence of contact terms after LSZ reduction on the non-displayed, on-shell legs. Note this absence only holds as long as $p \neq 0$. The just derived equation can be used to solve part of the lightcone gauge perturbation theory, at least at tree level. Here, it can be used to interchange an added $\varphi$ leg for an $e$ field leg. Hence, the analysis just presented directly applies to graphs including $\varphi$ fields. In fact, we have observed in examples that adding a $\varphi$ improves large $z$ behaviour. In conclusion, the Lagrangian in equation \eqref{eq:largezverts} generates a perturbation theory which, as far as checked, in the large $z$ limit displays the structure of  equation \eqref{eq:knowngravityshift}, even off-shell. Explicit checks were performed for up to Green's functions with up to $12$ gravitons.  

At order $z^0$ one can have additional vertex contributions which involve the vertices in the Lagrangian beyond those displayed in equation \eqref{eq:largezverts}. We have been unable to formulate a general argument for its scaling contribution; this seems to depend crucially on the structure of the vertex.

\subsubsection*{Propagator dependence}
Let us briefly study the propagator contribution at order $z$ in more detail. It will be useful to study first the result of contracting an off-shell field on a vertex with two $q$ vectors, in the large $z$ limit. We conjecture,
\begin{equation}\label{eq:propscontribconjecture}
\lim_{z\rightarrow \infty} q_{m} q_{n} \langle \frac{\partial}{\partial e_{mn}} \mathcal{L}_{n>2}, X \rangle \stackrel{?}{=}   \mathcal{O}\left(z^{-1} \right) 
\end{equation}
for on-shell fields in the set $X$. This is in effect the graviton current contracted with $q$'s.  Naively, this quantity scales as $z^0$ by power counting. To the extend it can be proven, this eliminates the order $z$ term from the momentum dependence of the propagator. 

The scaling of the propagator comes with two q's contracted into the left and right currents. Because of the gauge choice, most terms vanish or are sub-leading. The only non-vanishing vertex is generated by
\begin{equation}\label{eq:largezprops}
\lim_{z\rightarrow \infty} q^m q^n \frac{\partial}{\partial e_{mn}} V_n =  (1+\varphi)^2  q_{m} q_{n} \left(\partial^{m} e_{ab} \partial^{n} e_{cd} G^{ac} G^{db}  \right) + \mathcal{O}\left(z^{-1} \right),
\end{equation}
where the sub-leading terms for instance involve a contraction of $q$ with one of the BCFW shifted legs. In addition, there are vertices generated by the Lagrangian in equation \eqref{eq:largezverts}. Combined with index ordering this can be used to verify equation \eqref{eq:propscontribconjecture} for the first few contributions. The contributions with internal $\varphi$ legs can be traded for $e$ legs using the Schwinger-Dyson equations as above. 

\begin{figure}[!htb]
\centering 
\includegraphics[scale=0.5]{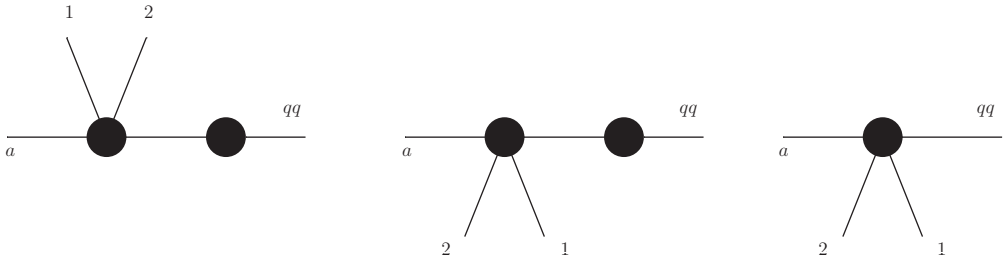}
\caption{\label{fig:i} Propagator large z contributions.}
\end{figure}

The simplest case involves two $q's$ contracted into a vertex with two $e$ fields in addition to the current leg contracted with the two $q's$. This current can be computed from three index-ordered Feynman graphs, see figure \ref{fig:i}, and is proportional to 
\begin{equation}
\propto (q \cdot p_1) (q \cdot p_a)+ \mathcal{O}\left(z^{-1} \right),
\end{equation}
which vanishes if the $a$ leg is taken to be on-shell. 

The next case is four $e$ fields. There are three index-ordered Feynman graphs which contribute: all legs on the vertex, or mixing two vertices from equations \eqref{eq:largezverts} and  \eqref{eq:largezprops}. Their sum vanishes up to sub-leading terms for on-shell leg $a$. Note that the other legs remain off-shell. For four additional $e$ fields the combinatorics is more complicated. Summing all graphs confirms \eqref{eq:propscontribconjecture}. Note that for this computation one can recycle the results obtained above as on-leg off-shell currents.  

\subsubsection*{Summarising results on sub-leading terms}

In this sub-sub-section we have explored the possibility to investigate sub-leading cancellations within Feynman graphs. The results above are enough to prove equation \eqref{eq:knowngravityshift} at order $z$ for tree level scattering amplitudes with up to eight legs. It is obvious that the above off-shell cancellations will drive even higher order cancellations and in general we have no doubt that the result equation \eqref{eq:knowngravityshift} follows for DFT.  These off-shell cancellations are very intriguing and deserve further study. Note that they seem closely related to improved BCFW shift behaviour of permutation sum shifts studied in \cite{Boels:2011mn}.

\section{Discussion}\label{sec:disc}

In this article we have studied the perturbation theory of the double field theory formulation of $\mathcal{N}=0$ supergravity: Einstein-Hilbert gravity coupled to a two-form and a dilaton. This is the low energy effective action of tree level closed string theory and is special as it features left-right index factorisation. Above we have verified explicitly that the DFT reproduces most the tree level S-matrix, with the exception of scattering amplitudes which do not involve any external gravitons. In principle these could be proven by pushing the analysis of BCFW shifts one order further than done above. Our result on large BCFW shifts for gravity show that gravity integrands can, in principle, be reconstructed from their single cuts. Obtaining explicit recursion relations at loop level would of course be even more exciting. 

A further interesting direction for future research is the issue of $\alpha'$ corrections to the double field theory. Above we showed that for four points / four field terms a fairly simple result may be obtained. In order to extend this, one should first obtain the completion of this result under gauge invariance. Then, one may study higher point amplitudes. The bottleneck here will be computational complexity; the lightcone gauge will be instrumental in such a program. On the other hand, given a result for  $\alpha'$ corrections obtained through for instance more geometric arguments, the lightcone gauge offers a quick way to verify consistency of the obtained result. This might have interesting cross-connections with work on DFT on curved background, \cite{Hohm:2015ugy}.

A prime motivation for the present article was color-kinematic duality. The hope was that by studying higher point amplitudes one would start to see some of the kinematic algebra structure needed to project the gravitational scattering amplitudes down to Yang-Mills scattering amplitudes. Although some of this structure makes an appearance in the large BCFW shift where it was shown to involve index ordering, in general we have been unable to identify a clear `square root'. In this sense, it seems color-kinematic duality and double field theory share some characteristics, but seem to address fundamentally different structures: we have found little evidence that color-kinematic duality and T-duality are equivalent beyond index factorisation. On the other hand, the double field theory equations of motion neatly bundle graviton, two-form and dilaton into a single package, which might be useful technically for studies of double copies beyond scattering amplitudes, e.g.  as in \cite{Monteiro:2014cda}. 

Finally, it might be interesting to integrate out the dependent degree of freedom out of the DFT action to obtain the Lightcone Lagrangian of the theory, along the lines of \cite{Goroff:1983hc} for Einstein-Hilbert. This should be an easier way to derive fully supersymmetric versions of DFT.

\acknowledgments
It is a pleasure to thank Tobias Hansen, Henrik Hanssen, Olaf Hohm,
Reinke Sven Isermann, and Charles Strickland-Constable for discussions. This work was
supported by the German Science Foundation (DFG) under the
Collaborative Research Center (SFB) 676 ``Particles, Strings and the
Early Universe''. Extensive tensor algebra computations were performed
using the Mathematica package FeynCalc \cite{Mertig:1990an}. Some Feynman diagrams were drawn
using the \LaTeX~package PGF/TikZ. Jaxodraw \cite{Binosi:2008ig} based on Axodraw \cite{Vermaseren:1994je} has been used to produce some of the figures.

\appendix
\section{Conventions}\label{a:indices} 
The metric signature is ``mostly plus''. This paper is about
perturbative quantum gravity around a constant background Minkowski
metric $G_{ij}=diag(-1,1,\ldots,1)$. The Mandelstam variables are
defined in terms of the outgoing momenta of the 4-point amplitude as 
\begin{equation}
  s = (k_1 + k_2)^2\, , \qquad t = (k_1+k_3)^2\, , \qquad
  u=(k_1+k_4)^2\, .
\end{equation}
More generally, beyond the 4-point amplitude we use
\begin{equation}
  s_{ij} = (k_i+k_j)^2\, , \qquad s_{ijk} = (k_i+k_j+k_k)^2\, , \qquad
  \ldots\, .
\end{equation}

\section{Physical degrees of freedom in DFT: equations of motion}\label{a:lightcone}
We find it reassuring to check that the DFT Lagrangian describes
precisely the physical degrees of freedom associated to the massless
excitations of the closed bosonic string. To this end, we analyse the
linearised equations of motion in the light-cone gauge.

From the kinetic terms \eqref{DFT_kineticterms} in the DFT Lagrangian
without a gauge-fixing term one obtains the following linearised
equations of motion:
\begin{align}\label{a:linearisedesom}
  \partial^2\varphi - \frac{1}{4}\partial^i\partial^j e_{ij}
  &=0\, ,\nonumber\\
  \partial^2 e_{ij} - \partial_i \partial^k e_{kj}
  - \partial_j \partial^k e_{ik} + 4\partial_i \partial_j \varphi &
  =0\, .
\end{align}
The full DFT Lagrangian has a gauge symmetry. For restricted fields
that do not depend on the coordinates $\tilde{x}^i$ the gauge
transformations are given in terms of gauge parameters $\lambda^i(x)$
and $\bar{\lambda}^i(x)$ as follows \cite{Hull:2009zb,Hohm:2011dz}:
\begin{align}\label{a:gaugetransformations}
  \delta_{\lambda,\bar{\lambda}} e_{ij} = & \partial_i \bar{\lambda}_j
  + \partial_j \lambda_i \nonumber\\
  &+
  \frac{1}{2}(\lambda+\bar{\lambda})\cdot\partial e_{ij} +
  \frac{1}{2}(\partial_j \bar{\lambda}^k - \partial^k \bar{\lambda}_j)
  e_{ik} + \frac{1}{2} (\partial_i \lambda^k - \partial^k \lambda_i)
  e_{kj}\nonumber\\
  & -\frac{1}{4} e_{ik} (\partial^l \bar{\lambda}^k+ \partial^k
  \lambda^l) e_{lj}\, ,\nonumber\\
  \delta_{\lambda,\bar{\lambda}} \varphi = &
  \frac{(1+\varphi)}{4} \partial\cdot (\lambda+\bar{\lambda}) +
  \frac{1}{2}(\lambda+\bar{\lambda})\cdot \partial\varphi\, .
\end{align}
Here, the gauge transformation of $\varphi$ has been derived from
\begin{equation}
  \delta_{\lambda,\bar{\lambda}} d =
  -\frac{1}{4} \partial\cdot(\lambda+\bar{\lambda}) +
  \frac{1}{2}(\lambda + \bar{\lambda})\cdot \partial d\, ,
\end{equation}
using the redefinition $1+\varphi = e^{-d}$ as in
\cite{Hohm:2011dz}. It can easily be checked that the linearised field
equations \eqref{a:linearisedesom} are invariant under
\eqref{a:gaugetransformations} (modulo higher orders of fields). 

In order to extract the physical degrees of freedom of a given field
excitation, the fields $e_{ij}$ and $\varphi$ are Fourier-transformed
and are assumed to describe infinitesimal fluctuations,
$\mathcal{O}\left(\frac{1}{\infty}\right)$ say, around the
background. As a result, in terms of light-cone components,
\begin{equation}
  x^+ = \frac{1}{\sqrt{2}} (x^0+x^1)\, ,\qquad
  x^- = \frac{1}{\sqrt{2}} (x^0-x^1)\, ,
\end{equation}
one finds
\begin{align}
  \delta_{\lambda,\bar{\lambda}} e^{++} & = ip^+(\lambda^+ +
  \bar{\lambda}^+) + \mathcal{O}\left ((\lambda,\bar{\lambda},e)^2
  \right) \, ,\nonumber\\
  \delta_{\lambda,\bar{\lambda}} e^{+-} & = ip^+ \bar{\lambda}^- +
  ip^- \lambda^+ + \mathcal{O}\left ((\lambda,\bar{\lambda},e)^2
  \right) \, ,\nonumber\\
  \delta_{\lambda,\bar{\lambda}} e^{-+} & = ip^- \bar{\lambda}^+ +
  ip^+ \lambda^- + \mathcal{O}\left ((\lambda,\bar{\lambda},e)^2
  \right) \, ,\nonumber\\
  \delta_{\lambda,\bar{\lambda}} e^{+I} & = ip^+ \bar{\lambda}^I +
  ip^I \lambda^+ + \mathcal{O}\left ((\lambda,\bar{\lambda},e)^2
  \right) \, ,\nonumber\\
  \delta_{\lambda,\bar{\lambda}} e^{I+} & = ip^I \bar{\lambda}^+ +
  ip^+ \lambda^I + \mathcal{O}\left ((\lambda,\bar{\lambda},e)^2
  \right) \, ,
\end{align}
where $I\in\{2,\ldots, D-1 \}$ label the transverse directions. For
physical excitations with $p^+\neq 0$ one can therefore successively
gauge away all the infinitesimal plus components,
\begin{equation}
  e^{+i} = 0 + \mathcal{O}\left(\frac{1}{\infty^2}\right)\, , \qquad 
  e^{i+} = 0 + \mathcal{O}\left(\frac{1}{\infty^2}\right)\, ,
\end{equation}
for all $i$. Using the linearised equations of motion
\eqref{a:linearisedesom}, one then finds
\begin{equation}
  \varphi = 0 + \mathcal{O}\left(\frac{1}{\infty^2}\right)\, , \qquad p_i
  e^{ij} = 0 + \mathcal{O}\left(\frac{1}{\infty^2}\right)\, , \qquad 
  p_j e^{ij} = 0+ \mathcal{O}\left(\frac{1}{\infty^2}\right)\, ,
\end{equation}
which eliminates $\varphi$ and shows that components $e_{-i}$, $e_{i-}$
are given in terms of the physical degrees of freedom $e_{IJ}$
satisfying the Klein-Gordon equation
\begin{equation}
  p^2 e^{IJ} = 0+ \mathcal{O}\left(\frac{1}{\infty^2}\right)\, ,
\end{equation}
for all $I,J$. Being an $SO(D-2)$-tensor, $e^{IJ}$ precisely encodes
the massless excitations of the closed bosonic string. 

A natural question to ask is whether or not one can gauge away
$\varphi(x)$ at each spacetime point simultaneously. In order to do so
one would have to solve the differential equation for
$\delta_{\lambda,\bar{\lambda}} \varphi(x) = -\varphi(x)$ for an
arbitrary (not necessarily infinitesimal) configuration
$\varphi(x)$. As of now, we are not sure if this is possible, since we
are unable to solve the differential equation resulting from
\eqref{a:gaugetransformations}.

We end this discussion by giving the field redefinitions to lowest
order that follow from the definitions in \cite{Hohm:2011dz}:
\begin{align}
  \phi &= \frac{1}{4} \text{Tr} (eG) - \varphi +
  \mathcal{O}(e^2)+\mathcal{O}(\varphi^2)\, , \nonumber\\
  \check{e} &= e + \mathcal{O}(e^2)\, ,
\end{align}
where $\phi$ is the dilaton field and $\check{e}$ is the finite
fluctuation around a constant background $G+B$ to be quantised
using the Lagrangian \eqref{lowenergyclosedstringaction}. Using
\eqref{a:gaugetransformations} it is easy to check that $\phi$ is in
fact gauge-invariant to lowest order (and beyond). Note that the
dilaton $\phi=\frac{1}{4} {e_i}^i-\varphi +\ldots$ contracts left and
right indices and, hence, the physical degree of the dilaton is
obscured in the factorised Lagrangian. This is the price we have to
pay for the factorisation property. 

\section{Derivation of the $q$-light-cone propagator: canonical approach}\label{app:lightcone}
Generalising the discussion in \eqref{a:lightcone} choose $q$-light-cone gauge
\begin{equation}\label{a:qgauge}
  q^i e_{ij}(p) = 0\, , \qquad q^j e_{ij}(p)=0\, ,
\end{equation}
for excitations with $p\cdot q\neq 0$ and a given light-like Lorentz
vector $q^i$. From the equations of motion one again obtains
$\varphi=0$. In Yang-Mills theory this gauge $(q^i A_i(p)=0)$ can be
used to derive $q$-transversal propagators which can be useful in the 
discussion of the large-$z$ behaviour (e.g.\cite{Boels:2010nw}). In gravity an obstacle to a straightforward derivation of the propagators exists which may be interesting to others.

The standard, usually fail-safe method of deriving propagators adds generic sources $J_{ij}$ and $J$ to the DFT
fields, 
\begin{equation}\label{a:L2Jsources}
  \mathcal{L}^{(2)'} = \mathcal{L}^{(2)}+ J_{ij}e^{ij}+J \varphi
\end{equation}
where the quadratic Lagrangian $\mathcal{L}^{(2)}$ without
gauge-fixing is given in \eqref{DFT_kineticterms}. Solving the classical equation of motion and plugging this back into the Lagrangian gives a Lagrangian quadratic in both fields and in sources: the second part contains the propagators.

It is because of $q^2=0$ that there are only finitely many terms in the most general
linear shifts  
\begin{align}
  e_{ij} = & e_{ij}' + e_{ijst} J^{st} + \tilde{e}_{ij} J\,
  ,\nonumber\\
  \varphi = & \varphi' + \varphi_{st} J^{st} + \tilde{\varphi} J
\end{align}
where schematically
\begin{align}
  e_{ijst} = &  \frac{1}{p^2} \Big ( G\, G + \frac{G\, p\, q}{p\cdot q}
  + \frac{G\,q\,q\,p^2}{(p\cdot q)^2}+\frac{G\,p\,p}{p^2} +
  \frac{p\,p\,p\,p}{p^4} \nonumber\\
   &\qquad+ \frac{p\,p\,p\,q}{p\cdot q p^2} + \frac{p\,p\,q\,q}{(p\cdot
    q)^2} + \frac{p\,q\,q\,q\,p^2}{(p\cdot q)^3} +
  \frac{q\,q\,q\,q\,p^4}{(p\cdot q)^4} \Big)_{ijst}\, , \nonumber\\
  \tilde{e}_{ij} = &\frac{1}{p^2} \Big ( G + \frac{p\, q}{p\cdot q} +
  \frac{p\, p}{p^2} \Big)_{ij}\, , \nonumber\\
  \varphi_{ij} = &\frac{1}{p^2} \Big ( G + \frac{p\, q}{p\cdot q} +
  \frac{p\, p}{p^2} \Big)_{ij}\, , \nonumber\\
  \tilde{\varphi}  = &\frac{1}{p^2}\, .
\end{align}
In order for $e_{ij}'$ to also satisfy the gauge conditions
\eqref{a:qgauge} one also requires
\begin{align}
  q^i e_{ijst} = q^j e_{ijst} =0\,, \qquad q^i \tilde{e}_{ij} = q^j
  \tilde{e}_{ij} = 0\, .
\end{align}
One finds that one cannot further constrain the coefficients such that in the $q$-light-cone gauge all mixed terms in
\eqref{a:L2Jsources} vanish. As a consequence, in this gauge it is impossible to derive the propagator  of DFT using this method. The same problem arises if one
does not introduce a current for $\varphi$ (i.e.\ $J=0$). On the other hand, note that upon restricting $J^{ij}$ to $J^{ij}q_i q_j=0$ it is 
possible to get rid of the mixed terms. This, however, comes at the cost of introducing ambiguities in the quadratic term that lead to
ambiguities in the propagator. In the main text a work-around is shown. 

It is easy to check that the current method gives the well-known result in Yang-Mills theory. However we also studied $q$-light-cone gauge in the
original  formulation of $\mathcal{N}=0$ supergravity given in
\eqref{lowenergyclosedstringaction}. The propagator of the dilaton
being gauge-invariant is clearly not affected by the gauge choice. The
kinetic terms for the $B$-field read \cite{Hohm:2010jy}
\begin{equation}
  -\frac{1}{12} H^2 = -\frac{1}{4} \left (G^{ac}G^{bd} G^{xy} - 2
    G^{ay} G^{bd} G^{cx} \right) \partial_x b_{ab} \partial_y b_{cd}\,
  . 
\end{equation}
Here, in $q$-light-cone gauge 
\begin{equation}
  q^a b_{ab} = q^b b_{ab}=0
\end{equation}
there is a linear shift
\begin{equation}
  b_{ab} = b_{ab}' + \omega_{abcd} J^{cd}
\end{equation}
for a generic, antisymmetric source $J^{ab}$ that respects the gauge
condition and removes the mixed terms in the Lagrangian. The resulting
propagator is unique and $q$-transversal:{\tiny
\begin{align}
  \frac{i}{p^2} \Big [ \left (G^{ms} G^{nt} - G^{mt} G^{ns} \right ) +
  \frac{G^{ns} \left(q^mp^t+p^mq^t \right) - G^{nt} \left (q^m p^s + p^m q^s
    \right) + G^{mt} \left (q^n p^s+p^nq^s \right)- G^{ms} \left
      (q^np^t+p^nq^t \right)}{p\cdot q}\nonumber\\
+\frac{p^mq^n p^t q^s-q^mp^np^tq^s+q^m p^n p^s
  q^t-p^mq^np^sq^t}{(p\cdot q)^2} \Big ]
\end{align}}
As to the metric, expanding the Einstein-Hilbert
term in \eqref{lowenergyclosedstringaction} one finds (e.g.\
\cite{'tHooft:1974bx}: 
\begin{equation}\label{eq:einsteinhilbertquad}
  \sqrt{g} R|^{(2)} = h_{kl,x} h_{mn,y} \frac{1}{2} \left (G^{mn}
    G^{ly} G^{kx} - G^{mx} G^{ky} G^{ln} + \frac{1}{2} G^{xy}G^{km}
    G^{ln} - \frac{1}{2} G^{xy} G^{kl} G^{mn} \right)
\end{equation}
It turns out that no shift of $h_{ij}$ exists that eliminates the
mixed terms for generic symmetric sources $J_{ij}$. The obstacle to direct derivation  of a $q$-light-cone propagator in DFT is thus related to its
impossibility in pure gravity. Note that again for restricted sources $J_{ij}$ with $J_{ij} q^iq^j$ mixed terms can be eliminated which, however, leads to ambiguities in the propagator. 

\subsection{Lightcone gauge propagator in Einstein-Hilbert}
For completeness we will derive the lightcone gauge propagator in Einstein-Hilbert gravity using the method indicated in the text. For this, expand the symmetry tensor $h$ in lightcone gauge (any contraction with $q$ vanishes as
\begin{equation}
h_{mn} = \tilde{h}_{mn} + \frac{h_{m} q_{n} + h_{n} q_{m} } {q\cdot p}+  \frac{h  q_{m} q_n } { (q\cdot p)^2}
\end{equation}
where
\begin{align}
\tilde{h}_{mn} & =  R_{m}{}^i R_{n}{}^j  h_{ij} \\
h_{m} & = p^{i} R_{m}{}^j h_{jm}\\
h & = p^{i} p^j h_{ij}
\end{align}
Plugging this expansion into equation \eqref{eq:einsteinhilbertquad} gives after the dust settles
\begin{equation}
 \sqrt{g} R|^{(2)}  = \frac{1}{2} \tilde{h}_{ij} p^2  \tilde{h}^{ij}  - \frac{1}{2}  \left(\tilde{h}_{i}{}^i \right)^2 + h   \left(\tilde{h}_{i}{}^i \right) - h_m h^m
 \end{equation}
Isolating the trace part of the symmetric tensor as
\begin{equation}
\tilde{h}_{ij} = \hat{h}_{ij} + \frac{1}{D-2} R_{ij}  \tilde{h}_{m}{}^m 
\end{equation}
shows that the degrees of freedom neatly split into traceless symmetric tensor $\hat{h}_{ij}$, the trace of $\tilde{h}_{ij}$, the vector $h_{m}$ and the scalar $h$, 
\begin{equation}
 \sqrt{g} R|^{(2)}  = \frac{1}{2} \hat{h}_{ij} p^2  \hat{h}^{ij}  - \frac{(D-3)}{2 (D-2)}  \left(\tilde{h}_{i}{}^i \right)^2 + h   \left(\tilde{h}_{i}{}^i \right) - h_m h^m
\end{equation}
The kinetic mixing term between trace and scalar $h$ is important, as introducing sources $K h + J \tilde{h}_{i}{}^i$ and inverting gives for the sources
\begin{equation}
\sim \frac{3-d}{2 (d-2)} K^2 p^2 - J K
\end{equation}
Importantly, the quadratic Lagrangian only generates $hh$ and $h  \tilde{h}_{i}{}^i$ correlators which do not have poles,
\begin{align}
\langle h \, h \rangle & =  \ii \frac{3-d}{(d-2)} p^2  \\  
\langle h  \tilde{h}_{i}{}^i \rangle & = - \ii
\end{align}
Hence, these degrees of freedom are auxilliary. Similarly, for the field $h_m$ we obtain
\begin{equation}
\langle h_m  h_k \rangle  = - \ii 2 R_{mk}
\end{equation}
while for the traceless symmetric tensor
\begin{equation}
\langle \hat{h}_{ij} \hat{h}_{mn} \rangle =\ii \frac{R_{im} R_{jn} }{p^2}
\end{equation}
holds. This completes the list of non-vanishing correlators in the lightcone gauge. The only correlator with a pole is the last one. Hence the physical degrees of freedom of the gravitational field, in lightcone gauge, are contained in the traceless symmetric tensor $\hat{h}_{ij}$ which is transverse and orthogonal to $q$. This is of course well-known physical field content of Einstein gravity.

Plugging the expansion into the correlator of the general field $h_{ij}$ gives for the full correlator
\begin{multline}
\langle h_{mn} h_{kl} \rangle = \ii \left( \frac{R_{mk} R_{nl} }{p^2}  \right) +  \\
\frac{2 \ii }{(q\cdot p)^2} \left( q_{m} q_{k} R_{nl}+  q_{m} q_{l} R_{nk}+  q_{n} q_{k} R_{nl}+ q_{n} q_{l} R_{mk} \right) \\ + \frac{\ii}{(q\cdot p)^4}  \left( q_{m} q_{n} q_{k} q_{l}  \frac{3-d}{(d-2)} p^2  \right) -
\frac{\ii}{(q\cdot p)^2}  \left(\frac{R_{mn}}{d-2} q_{k} q_{l} + \frac{R_{kl}}{d-2} q_{m} q_{n}\right)
\end{multline}
This is the lightcone-gauge propagator of Einstein gravity. Plugging in the definition of the $R$ projector does not yield particularly insightful results. In particular, the numerator does not have the form of the numerator squared of the Yang-Mills lightcone gauge propagator.

\section{Partial amplitudes in Yang-Mills theory}\label{a:partialamplitudes} 
In Yang-Mills theory partial Yang-Mills amplitudes at tree-level are
defined by 
\begin{equation}
  \mathcal{A}^{\text{tree}}_n = g^{n-2} \sum_{\sigma\in P_3/Z_3} \text{Tr}
  \left(T^{a_{\sigma(1)}} \ldots T^{a_{\sigma(n)}} \right)\,
  A_n^{\text{tree}}(\sigma(1),\ldots,\sigma(n))\, ,
\end{equation}
where $g$ is the $SU(N)$ gauge coupling and the sum is over non-cyclic
permutations. The gauge group generators are chosen to satisfy
\begin{equation}
  \text{Tr}\left (T^A T^B \right) = \frac{1}{2} \delta^{AB}\, ,\qquad
  [T^A, T^B] = i f^{ABC} T^C\, ,
\end{equation}
and the $SU(N)$ Fierz identity needed to express product of traces in
terms of sums of single traces of generators is
\begin{equation}
  T^{Aij} T^{Akl} = \frac{1}{2} \left (\delta^{il} \delta^{jk} -
    \frac{1}{N} \delta^{ij} \delta^{kl} \right)\, .
\end{equation}
The partial 3-point amplitude reads
\begin{equation}
  \left(\mathcal{A}_3^{\text{
      partial}}(1,2,3)\right)^{ace} = 2i \left (G^{ae}(k_1-k_3)^c +
  G^{ce}(k_3-k_2)^a + G^{ac} (k_2-k_1)^e \right)\, ,
\end{equation}
and the partial 4-point amplitudes are given by
\begin{eqnarray}
  \left (\mathcal{A}_4^{\text{partial}}(1,2,3,4)\right)^{aceg}
  & = & 2i(2G^{ae}G^{cg} - G^{ag} G^{ce} - G^{ac} G^{eg})\nonumber\\
  & &+ \frac{2i}{s} \left (2G^{ai} k_1^c - 2G^{ci} k_2^a +
    G^{ac}(k_2-k_1)^i \right)
  \left (2\delta_i^e k_3^g-2\delta_i^g k_4^e+G^{eg}(k_4-k_3)_i \right)
  \nonumber\\
  & &+ \frac{2i}{u} \left (2G^{ai} k_1^g - 2 G^{gi} k_4^a +
    G^{ag}(k_4-k_1)^i \right) \left (2\delta_i^e k_3^c - 2 \delta_i^c
    k_2^e + G^{ce} (k_2-k_3)_i \right)\, ,\nonumber\\
  \left (\mathcal{A}_4^{\text{partial}}(1,2,4,3)\right)^{acge} &
  = & \% \text{ with } \{u\leftrightarrow t, k_3\leftrightarrow k_4,
  e\leftrightarrow g\}\, ,\nonumber\\
  \left (\mathcal{A}_4^{\text{partial}}(1,3,2,4)\right)^{aecg} &
  = & \% \text{ with } \{s\leftrightarrow t, k_2\leftrightarrow k_3,
  c\leftrightarrow e\}\, .
\end{eqnarray}
The partial amplitudes satisfy the Ward identities
\begin{equation}
  \left(\mathcal{A}_3^{\text{partial}}(1,2,3) \right)^{ace}
  k_{1a} \, \xi_c(k_2)\, \xi_e(k_3) = 0\, , \text{ etc.}
\end{equation}
(and analogous expressions for $n$ points) due to momentum conservation
and transversality $\xi(k)\cdot k =0$ of the polarisation vectors.  


\bibliographystyle{JHEP}

\bibliography{myrefs2}

\providecommand{\href}[2]{#2}\begingroup\raggedright\begin{thebibliography}{10}

\bibitem{'tHooft:1974bx}
G.~'t~Hooft and M.~Veltman, {\it {One loop divergencies in the theory of
  gravitation}},  {\em Annales Poincare Phys.Theor.} {\bf A20} (1974) 69--94.

\bibitem{Goroff:1985th}
M.~H. Goroff and A.~Sagnotti, {\it {The Ultraviolet Behavior of Einstein
  Gravity}},  {\em Nucl. Phys.} {\bf B266} (1986) 709.

\bibitem{vandeVen:1991gw}
A.~van~de Ven, {\it {Two loop quantum gravity}},  {\em Nucl.Phys.} {\bf B378}
  (1992) 309--366.

\bibitem{Bern:2015xsa}
Z.~Bern, C.~Cheung, H.-H. Chi, S.~Davies, L.~Dixon, and J.~Nohle, {\it
  {Evanescent Effects Can Alter Ultraviolet Divergences in Quantum Gravity
  without Physical Consequences}},
  \href{http://xxx.lanl.gov/abs/1507.0611}{{\tt arXiv:1507.0611}}.

\bibitem{Bern:2007hh}
Z.~Bern, J.~J. Carrasco, L.~J. Dixon, H.~Johansson, D.~A. Kosower, and
  R.~Roiban, {\it {Three-Loop Superfiniteness of N=8 Supergravity}},  {\em
  Phys. Rev. Lett.} {\bf 98} (2007) 161303,
  [\href{http://xxx.lanl.gov/abs/hep-th/0702112}{{\tt hep-th/0702112}}].

\bibitem{Beisert:2010jx}
N.~Beisert, H.~Elvang, D.~Z. Freedman, M.~Kiermaier, A.~Morales, and
  S.~Stieberger, {\it {E7(7) constraints on counterterms in N=8 supergravity}},
   {\em Phys. Lett.} {\bf B694} (2011) 265--271,
  [\href{http://xxx.lanl.gov/abs/1009.1643}{{\tt arXiv:1009.1643}}].

\bibitem{Bern:2008qj}
Z.~Bern, J.~J.~M. Carrasco, and H.~Johansson, {\it {New Relations for
  Gauge-Theory Amplitudes}},  {\em Phys. Rev.} {\bf D78} (2008) 085011,
  [\href{http://xxx.lanl.gov/abs/0805.3993}{{\tt arXiv:0805.3993}}].

\bibitem{Bern:2010ue}
Z.~Bern, J.~J.~M. Carrasco, and H.~Johansson, {\it {Perturbative Quantum
  Gravity as a Double Copy of Gauge Theory}},  {\em Phys.Rev.Lett.} {\bf 105}
  (2010) 061602, [\href{http://xxx.lanl.gov/abs/1004.0476}{{\tt
  arXiv:1004.0476}}].

\bibitem{Kawai:1985xq}
H.~Kawai, D.~Lewellen, and S.~Tye, {\it {A Relation Between Tree Amplitudes of
  Closed and Open Strings}},  {\em Nucl.Phys.} {\bf B269} (1986) 1.

\bibitem{Boels:2010nw}
R.~H. Boels, {\it {On BCFW shifts of integrands and integrals}},  {\em JHEP}
  {\bf 1011} (2010) 113, [\href{http://xxx.lanl.gov/abs/1008.3101}{{\tt
  arXiv:1008.3101}}].

\bibitem{Bork:2015zaa}
L.~V. Bork, D.~I. Kazakov, M.~V. Kompaniets, D.~M. Tolkachev, and D.~E.
  Vlasenko, {\it {Divergences in maximal supersymmetric Yang-Mills theories in
  diverse dimensions}},  \href{http://xxx.lanl.gov/abs/1508.0557}{{\tt
  arXiv:1508.0557}}.

\bibitem{Bern:1999ji}
Z.~Bern and A.~K. Grant, {\it {Perturbative gravity from QCD amplitudes}},
  {\em Phys.Lett.} {\bf B457} (1999) 23--32,
  [\href{http://xxx.lanl.gov/abs/hep-th/9904026}{{\tt hep-th/9904026}}].

\bibitem{Hohm:2011dz}
O.~Hohm, {\it {On factorizations in perturbative quantum gravity}},  {\em JHEP}
  {\bf 1104} (2011) 103, [\href{http://xxx.lanl.gov/abs/1103.0032}{{\tt
  arXiv:1103.0032}}].

\bibitem{Hohm:2010xe}
O.~Hohm and S.~K. Kwak, {\it {Frame-like Geometry of Double Field Theory}},
  {\em J. Phys.} {\bf A44} (2011) 085404,
  [\href{http://xxx.lanl.gov/abs/1011.4101}{{\tt arXiv:1011.4101}}].

\bibitem{Hohm:2013bwa}
O.~Hohm, D.~L{\"u}st, and B.~Zwiebach, {\it {The Spacetime of Double Field
  Theory: Review, Remarks, and Outlook}},  {\em Fortsch. Phys.} {\bf 61} (2013)
  926--966, [\href{http://xxx.lanl.gov/abs/1309.2977}{{\tt arXiv:1309.2977}}].

\bibitem{Hull:2009mi}
C.~Hull and B.~Zwiebach, {\it {Double Field Theory}},  {\em JHEP} {\bf 0909}
  (2009) 099, [\href{http://xxx.lanl.gov/abs/0904.4664}{{\tt
  arXiv:0904.4664}}].

\bibitem{Hohm:2010jy}
O.~Hohm, C.~Hull, and B.~Zwiebach, {\it {Background independent action for
  double field theory}},  {\em JHEP} {\bf 1007} (2010) 016,
  [\href{http://xxx.lanl.gov/abs/1003.5027}{{\tt arXiv:1003.5027}}].

\bibitem{Siegel:1993th}
W.~Siegel, {\it {Superspace duality in low-energy superstrings}},  {\em
  Phys.Rev.} {\bf D48} (1993) 2826--2837,
  [\href{http://xxx.lanl.gov/abs/hep-th/9305073}{{\tt hep-th/9305073}}].

\bibitem{Hohm:2014xsa}
O.~Hohm and B.~Zwiebach, {\it {Double field theory at order $\alpha'$}},  {\em
  JHEP} {\bf 11} (2014) 075, [\href{http://xxx.lanl.gov/abs/1407.3803}{{\tt
  arXiv:1407.3803}}].

\bibitem{Marques:2015vua}
D.~Marques and C.~A. Nunez, {\it {T-duality and ??-corrections}},  {\em JHEP}
  {\bf 10} (2015) 084, [\href{http://xxx.lanl.gov/abs/1507.0065}{{\tt
  arXiv:1507.0065}}].

\bibitem{Lee:2015kba}
K.~Lee, {\it {Quadratic ??-corrections to heterotic double field theory}},
  {\em Nucl. Phys.} {\bf B899} (2015) 594--616,
  [\href{http://xxx.lanl.gov/abs/1504.0014}{{\tt arXiv:1504.0014}}].

\bibitem{Gross:1986iv}
D.~J. Gross and E.~Witten, {\it {Superstring Modifications of Einstein's
  Equations}},  {\em Nucl. Phys.} {\bf B277} (1986) 1.

\bibitem{Boels:2012ie}
R.~H. Boels and D.~O'Connell, {\it {Simple superamplitudes in higher
  dimensions}},  {\em JHEP} {\bf 06} (2012) 163,
  [\href{http://xxx.lanl.gov/abs/1201.2653}{{\tt arXiv:1201.2653}}].

\bibitem{Boels:2013jua}
R.~H. Boels, {\it {On the field theory expansion of superstring five point
  amplitudes}},  {\em Nucl. Phys.} {\bf B876} (2013) 215--233,
  [\href{http://xxx.lanl.gov/abs/1304.7918}{{\tt arXiv:1304.7918}}].

\bibitem{Goroff:1983hc}
M.~Goroff and J.~H. Schwarz, {\it {$D$-dimensional Gravity in the Light Cone
  Gauge}},  {\em Phys. Lett.} {\bf B127} (1983) 61--64.

\bibitem{Leibbrandt:1987qv}
G.~Leibbrandt, {\it {Introduction to Noncovariant Gauges}},  {\em Rev. Mod.
  Phys.} {\bf 59} (1987) 1067.

\bibitem{ArkaniHamed:2008yf}
N.~Arkani-Hamed and J.~Kaplan, {\it {On Tree Amplitudes in Gauge Theory and
  Gravity}},  {\em JHEP} {\bf 0804} (2008) 076,
  [\href{http://xxx.lanl.gov/abs/0801.2385}{{\tt arXiv:0801.2385}}].

\bibitem{Britto:2004ap}
R.~Britto, F.~Cachazo, and B.~Feng, {\it {New recursion relations for tree
  amplitudes of gluons}},  {\em Nucl.Phys.} {\bf B715} (2005) 499--522,
  [\href{http://xxx.lanl.gov/abs/hep-th/0412308}{{\tt hep-th/0412308}}].

\bibitem{Britto:2005fq}
R.~Britto, F.~Cachazo, B.~Feng, and E.~Witten, {\it {Direct proof of tree-level
  recursion relation in Yang-Mills theory}},  {\em Phys.Rev.Lett.} {\bf 94}
  (2005) 181602, [\href{http://xxx.lanl.gov/abs/hep-th/0501052}{{\tt
  hep-th/0501052}}].

\bibitem{Boels:2011mn}
R.~H. Boels and R.~S. Isermann, {\it {Yang-Mills amplitude relations at loop
  level from non-adjacent BCFW shifts}},  {\em JHEP} {\bf 03} (2012) 051,
  [\href{http://xxx.lanl.gov/abs/1110.4462}{{\tt arXiv:1110.4462}}].

\bibitem{Hohm:2015ugy}
O.~Hohm and D.~Marques, {\it {Perturbative Double Field Theory on General
  Backgrounds}},  \href{http://xxx.lanl.gov/abs/1512.0265}{{\tt
  arXiv:1512.0265}}.

\bibitem{Monteiro:2014cda}
R.~Monteiro, D.~O'Connell, and C.~D. White, {\it {Black holes and the double
  copy}},  {\em JHEP} {\bf 12} (2014) 056,
  [\href{http://xxx.lanl.gov/abs/1410.0239}{{\tt arXiv:1410.0239}}].

\bibitem{Mertig:1990an}
R.~Mertig, M.~Bohm, and A.~Denner, {\it {FEYN CALC: Computer algebraic
  calculation of Feynman amplitudes}},  {\em Comput. Phys. Commun.} {\bf 64}
  (1991) 345--359.

\bibitem{Binosi:2008ig}
D.~Binosi, J.~Collins, C.~Kaufhold, and L.~Theussl, {\it {JaxoDraw: A graphical
  user interface for drawing Feynman diagrams. Version 2.0 release notes}},
  {\em Comput. Phys. Commun.} {\bf 180} (2009) 1709--1715,
  [\href{http://xxx.lanl.gov/abs/0811.4113}{{\tt arXiv:0811.4113}}].

\bibitem{Vermaseren:1994je}
J.~Vermaseren, {\it {Axodraw}},  {\em Comput.Phys.Commun.} {\bf 83} (1994)
  45--58.

\bibitem{Hull:2009zb}
C.~Hull and B.~Zwiebach, {\it {The Gauge algebra of double field theory and
  Courant brackets}},  {\em JHEP} {\bf 0909} (2009) 090,
  [\href{http://xxx.lanl.gov/abs/0908.1792}{{\tt arXiv:0908.1792}}].

\end{thebibliography}\endgroup

\end{document}